# How to Achieve the Best SRF Performance: (Practical) Limitations and Possible Solutions[1]

*C. Z. Antoine*
Cea Irfu, Centre d'Etudes de Saclay, 91191 Gif-sur Yvette Cedex, France

**Abstract**
This tutorial presents in the first part the requirement for the surface preparation of RF Niobium cavities and its justification in term s of physical origin of limitation (e.g. cleanliness and field emission, influence of the surface treatments, morphology and surface damage, etc.). in the second part we discuss the different models describing the ultimate limits of SRF cavities, and present one of the possible ways to overcome Niobium monopoly toward higher performances.

## 1 How to get a 'good' cavity

Whatever the application, a 'good' cavity is one that exhibits the highest possible accelerating field along with the highest possible quality factor. For a 1.3 GHz 'Tesla shape' cavity, that would be a quality factor in excess of $10^{10}$ and an accelerating gradient higher than 40 MV·m$^{-1}$, which corresponds to a surface magnetic field of ~170 mT.

The recipe for obtaining a good cavity is pretty well known, although up to the present time we still don't know why it works, and most of all, why it doesn't always work!

Basically, one needs to form high-purity, bulk niobium sheets to shape half-cells, then weld them together by electron beam melting, get rid of the internal surface damaged layer and any other surface impurities through (electro-) chemical polishing, and then finally bake the cavity at low temperature. In the next section, we will try to describe the origin of each of these specifications in the light of what we know about the physical and chemical properties of the material.

## 2 Field emission and particulate contamination

Field emission is an emission of electrons from a metal surface subjected to high electric fields (see Fig. 1). The phenomenon was studied by Fowler and Nordheim in 1928, and is attributed to the passage of electron tunnelling through the surface barrier.

Identified as a major cause of the limitation to the breakdown voltage of electrical devices due to the strains that it initiates, field emission has for decades been the object of many studies in d.c. RF studies started somewhat later.

---

[1] Most of the information in this tutorial is extracted, with permission, from "Materials and surface aspects in the development of SRF niobium cavities", EUCARD series on Accelerator Science Vol. XII (2011).

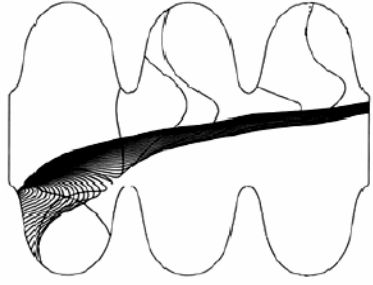

**Fig. 1:** The modelling of electron trajectories in a cavity

In accelerators, field emission is responsible for the unwanted absorption of RF power, dark current, and possible breakdown. In cavities, with fields of some tens of megavolts per metre, the measured emission current density is several orders of magnitude greater than predicted by the Fowler–Nordheim theory.

The observed current law follows a modified Fowler–Nordheim model (Eq. (14)), with two adjustable parameters: $Ae$, analogous to a surface, and $\beta$, a dimensionless parameter that corresponds to the enhancement factor of the electric field.[2] Typical values are $Ae \sim 10^{-15}$ m$^2$ and $\beta \sim 50–500$. $\Phi$ is the work function of the metal:

$$I_{DC}(E) = Ae \frac{1.54 \times 10^{-6} \beta^2 E^2}{\Phi} \exp\left(-\frac{6.83 \times 10^9 \Phi^{1,5}}{\beta E}\right). \tag{1}$$

Numerous experiments in d.c. and RF have shown a correlation between localized breakdown and field-emitting particles [1-16]. In particular, it has been shown that dust particles and scratches are good candidates for emission sites, and that the nature and the shape of the particles play a paramount role, independent of the substrate material.

Dust particles were quickly suspected, but it was apparently not possible to get a proper agreement with the Fowler–Nordheim law: their apex had a surface of some square micrometres ($10^{-12}$ m$^2$) and their $\beta_{apparent}$ was less than 10 (see Ref. [17]). Other influences have been explored, in particular the role of the oxide layer. Indeed, this has a very special 'amorphous-microcrystalline' structure with stacks of local defects, which can form ion- and/or electron-conducting channels that may possibly play an active role in the field emission (see, e.g., Refs. [18-21]).

However, it has been shown that increased thickness of the oxide layer – up to several hundreds of nanometres – due to anodizing does not change the nature or intensity of the field emission [22].

## 2.1 The influence of a particle's morphology on its emissivity

Specific devices inserted into the electron microscope allow us to identify *in situ* transmitter sites, and to obtain their image as well as their chemical composition, using *EDX* on samples. The field emission does indeed seem to be associated with the presence of particles or scratches on the surface. However, not all particles emit; it depends on their shape and size. An intentional surface contamination with particles of a similar kind but of different shape (see Fig. 2a) allows us to show that spherical particles, with a modest field increase factor ($\beta \sim 3$), emit far less than particles with more complex forms [23].

The presence of dust particles and roughness can locally increase the electric field. Roughness due to machining or etching does not play much of a role: direct topological evaluation of beta as seen in Section 3 and also as measured with a tunnel microscope by Niederman [17] shows that micron roughness induces betas of less than 2 to 10. Higher betas can only be attributed to the combination of defects at several size scales (see Figure 2b). Due to their fractal nature, most of the natural dust particles exhibit nanometric defects that add up to the general shape factor, and field enhancement factors of the order of several hundreds are easily achievable [10, 23, 24]. Moreover, in Ref. [25], it was shown that when the current density reaches $10^{11}$–$10^{12}$ A·m$^{-2}$, the emitting area (in the Fowler–Nordheim equation) decreases to a few square nanometres, a fact compatible with the melting of the emitter apex and the formation of a *Taylor cone*.

---

[2] Note that this electric field enhancement factor is similar to the magnetic field enhancement factor described in Section 3. They are both due to morphology (see below). A $\beta^{magn}$ of about 1.5–2.0 has dramatic consequences for the transition in the high magnetic field region, whereas the $\beta^{el}$ values observed in electron field emission are much higher (50–500) and cannot be related to surface roughness.

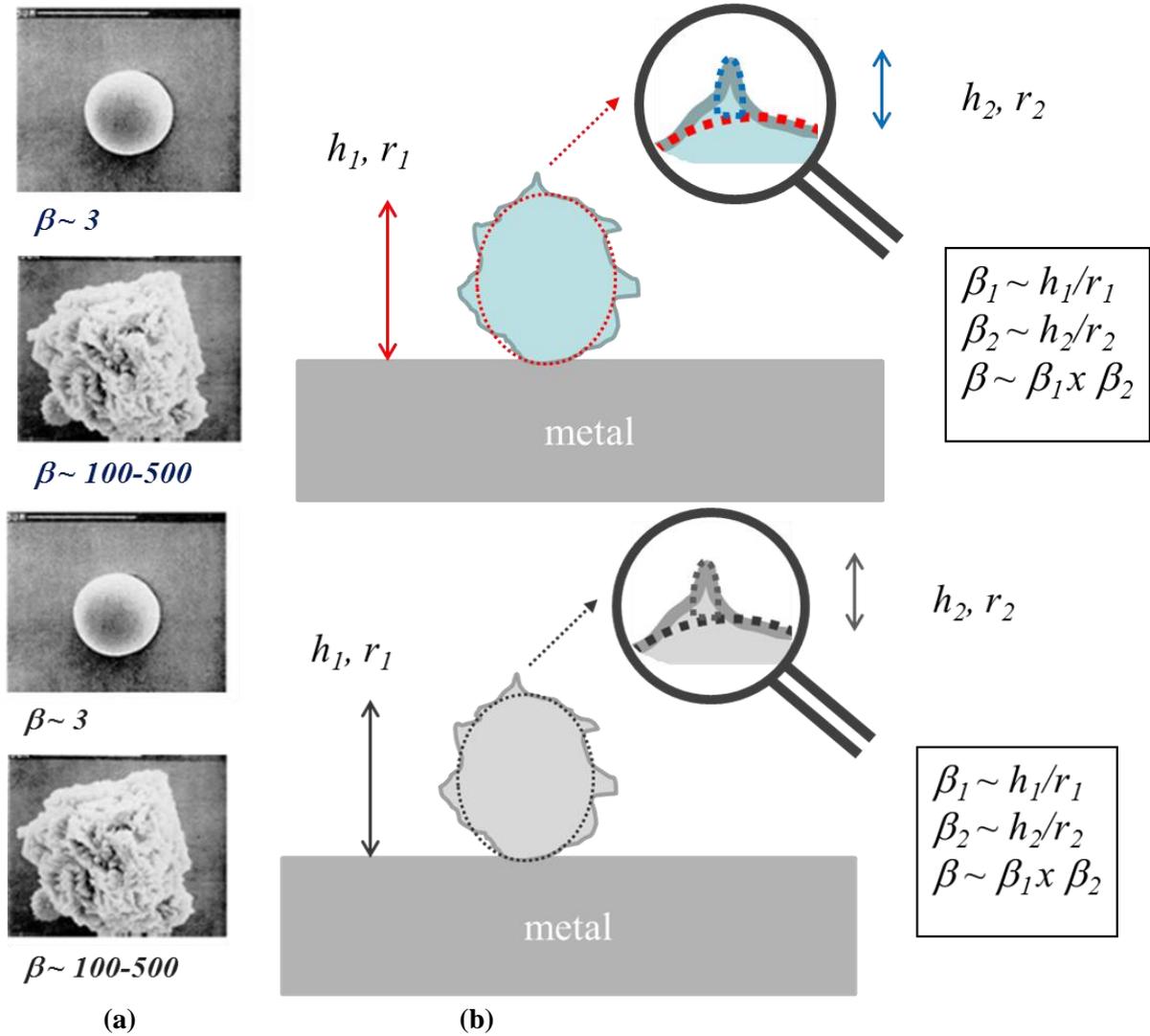

**Fig. 2:** (a) Nickel particles of varying morphology. (b) Modelling of the field increase factor due to the superposition of protuberances. This model allows us to understand why the perfectly spherical particle emits far less than the other.

Note the following about breakdown: in superconducting cavities, when breakdown occurs, it usually results in the formation of localized craters of a few microns, and it can either improve or decrease the cavity performance [9, 16]. Studies conducted on samples confirm this result, showing that processing is efficient only in 50% of the cases. In the other 50% of cases, breakdown results in a protruding particle welded on the surface, which is a very stable emitting site, and in a degradation of the cavity performance [9, 16]. With the improvement of cleaning techniques, surface fields up to 80–90 MV·m$^{-1}$ are currently observed without any breakdown or measurable field emission inside superconducting cavities [2], whereas breakdown is a common feature in normal conducting cavities.

The obvious difference between superconducting and normal conducting cavities is the operating temperature; a thermally activated phenomenon such as electromigration being negligible at low temperature [26].

## 2.2 Influence of assembly and vacuum handling

In the early 1990s, the dust created during cavity assembly was often cited as responsible for the field emission phenomena observed during the testing of cavities, but there were no formal proofs. The incidents were listed in the specifications of the experiments: leakage, problems during installation, change of seal, and so on. A statistical study of hundreds of test results stored at the laboratory showed that the threshold for the appearance of electrons was about 4–5 MV·m$^{-1}$ lower in the case of a problem during assembly. This effect can only be shown statistically: indeed, only a small portion of the surface may be at the origin of the emission, where the electric field is the strongest. The dust particles accumulate in a random way and are not necessarily localized in the 'danger zone'. Conversely, it will take just a single dust particle in that particular zone in order for the electrons to be emitted. Our work has been able to confirm, on a substantial basis, the intuitions that were generally held by the experimenters. This has led us to reconsider the role of dust particles [27]. Vacuum handling also calls for some precautions: in Ref. [28], general operations such as gasket assembly, pre-pumping and venting, and steady-state pumping have been monitored with a particle counter installed downstream inside the vacuum system. This shows that any human operation, shock or strong vibration liberates a lot of particles. Only steady-state pumping is rather benign, with the exception of start-up or possible arcing. Any valve manipulation generates particles if the pressure differential between the two sides is high.

## 3 What kind of niobium?

### 3.1 High-purity, bulk niobium

High purity is not required for superconductivity where a rather low mean free path leads to the lowest surface resistance. The surface resistance in RF is defined as a function of the temperature, as follows (see fig.3, left):

$$R_S = R_{BCS} + R_{res},\qquad(2)$$

where

$$R_{BCS} = A(\lambda_L^4, \xi_F, \ell, \sqrt{\rho_n})\frac{\omega^2}{T}e^{-\Delta/kT}.\qquad(3)$$

Here, $A$ is a constant depending on $\lambda_L$ (the penetration depth of the London field), $\xi$ (the coherence length of the Cooper pairs), $\ell$ (the mean free path of the quasi-particles), and $\rho_n$ (conductivity in the normal state); $\omega$ is the RF frequency and $\Delta$ is the superconducting gap. There exists a component $R_{res}$, which does not depend on the temperature. Its origin is not very clear, though it seems to be related to the conductivity of the material in the normal state, $\rho_n$. $R_{BCS}$ is due to the scattering of the remainder of the normal electrons of the superconductor over the lattice (where 'BCS' refers to Bardeen–Cooper–Schrieffer theory; see Section 4.1).

A high Residual Resistivity Ratio (RRR) is required for thermal stability rather than for good superconducting properties. As can be seen on fig. 3, centre, most of the bulk Nb cavities, with RRR~300, exhibit a higher $R_{BCS}$ than thin film Nb cavities with RRR ~15-30. On fig. 3, right, one shows that in presence of defects, the quench field is higher with higher RRR.

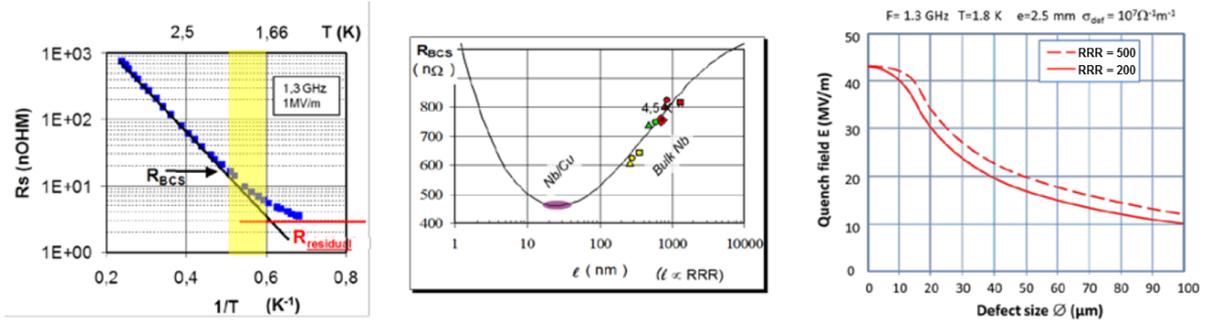

**Fig. 3:** From left to right: a typical surface resistance for a 1.3 GHz cavity at low field; the influence of the mean free path l on the BCS component of the surface resistance; and a thermal calculation showing the stabilization effect of a high RRR (after [29, 30]).

In the 3–15 K range, there is a direct relationship between the RRR and thermal conductivity [31]. At 4.2 K, the thermal conductivity from niobium is roughly equal to RRR/4 [2]. A superconductor is intrinsically a bad thermal conductor, as some of the (electrical and thermal) conduction electrons are paired into Cooper pairs and thus can no longer contribute to the heat transfer. To improve the thermal conductivity, it is essential to get rid of the main scattering sources; that is, interstitial light elements in the metal matrix. At lower temperatures ($\leq 2$ K), the major conduction mechanism is not related to electron propagation, but to phonon propagation. In such situations, the scattering sources are rather crystalline defects and thus fully recrystallized samples exhibit a large phonon peak, even with a rather low RRR. Examples of thermal conductivity curves vs RRR is given in Fig.4.

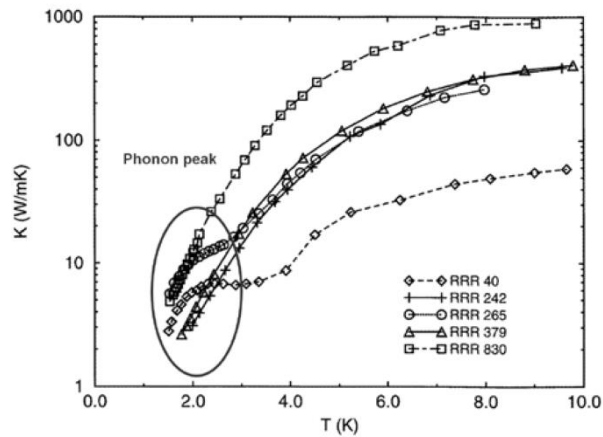

**Fig. 4:** The thermal conductivity of various RRR samples (after [31]). For well-recrystallized samples or monocrystals, the phonon peak can reach several tens of $W \cdot m^{-1} \cdot K^{-1}$ between 1.5 and 2 K, almost independent of the RRR.

Sputtered thin films, on the other hand, usually have a high $Q_0$ value at low field because of their lower mean free path, but they exhibit an early $Q$-slope, which confines them to low-field applications (see Fig.5). We will see in (§3.6) t that this situation is now changing.

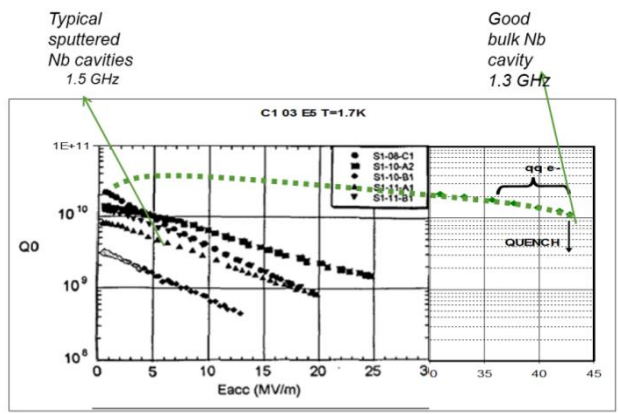

**Fig. 5:** A comparison between typical curves of niobium-sputtered films and bulk Nb cavities

## 3.2 Surface treatments

Empirically, an abrasion of at least 100–200 μm of the surface is necessary in order to obtain optimal cavity performance (see the discussion on the damaged layer below). Two main treatments exist, namely Buffered Chemical Polishing (BCP) and ElectroPolishing (EP) (see Fig.6). EP has supplanted BCP somewhat, at least in the hope of reaching record accelerating gradients, but is more difficult to perform reproducibly. A huge quantity of information is available in the literature about the optimization of each process, as well as on the research into alternative 'recipes'. The reader is invited to consult Superconducting Radio Frequency (SRF) tutorials for details. In this chapter, we will concentrate on alternative routes that seem to be the most promising in terms of cost.

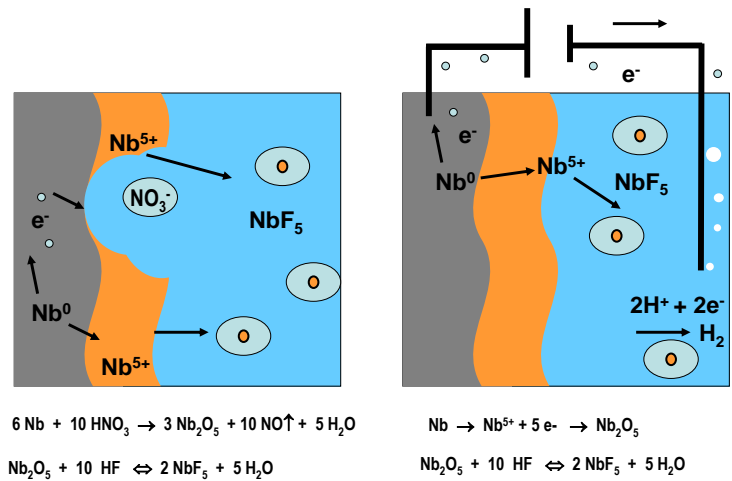

**Fig. 6:** A comparison between chemical polishing (left) and electropolishing (right). In both cases, niobium is oxidized into $Nb^{5+}$. In the case of chemical polishing, oxidation occurs because of the presence of a strong oxidant ($NO_3^-$) in the solution, while in electropolishing oxidation occurs because of the bias applied to the anode. Because of the presence of water, the stable form of Nb is $Nb_2O_5$; but HF decomposes the oxides into $NbF_5$, which is soluble in the solution.

Indeed, the removal of some hundreds of microns by electropolishing is costly and presents reproducibility issues. Chemical polishing leads to a detrimental roughness and in order to get a 'smooth' surface it is necessary to electropolish again by about 100 μm. Something similar happens when we apply industrial mechanical polishing techniques such as *tumbling*. The industrial process applied until recently left a damaged layer of about 100 μm, as can be inferred by the necessity to further etch the surface (by EP or BCP) [32-34]. It would be more interesting to find a reliable method for the first 1–200 μm and then finish the treatment with a light electropolishing operation (10–20 μm).

Very recently, the tumbling process was adapted in order to reduce the damaged layer, by including some of the features of the mechanical polishing applied in metallography and microscopy, which is known to produce a reduced damaged layer.

This chapter summarizes the main features concerning chemical and electrochemical treatments of the cavity surface. Basically, the (electro-)chemical reactions are the same, but the localization of the electrochemical cell is different: in chemical polishing, the oxidation/reduction occurs due to local differences of potential; while in EP, the potential is applied externally.

### 3.2.1 Buffered Chemical Polishing (BCP)

*Composition*
- ~2 Volumes of $H_3PO_4$ (buffer, very viscous).

- ~1 Volumes of $HNO_3$ (oxidant, transforms $Nb^0$ into $Nb^{5+}$).
- ~1 Volumes of HF (complexant of $Nb^{5+}$, dissolves the oxide layer formed by $HNO_3$ into $NbF_5$).
- Variation of composition allows us to adjust the etching rate.

*Pros*
- Easy to handle, middle stirring is necessary.
- Fast etching rate.
- Very reproducible.

*Cons*
- This is not 'polishing', but 'etching': all crystalline defects are preferentially attacked (etching pits, etching figures).
- Grains with various orientations are not etched at the same rate, which induces roughness!
- Except for a few cases, $E_{acc}^{max}$ ~ 25–30 MV·m$^{-1}$.

*Caution!*
- Do not process at temperatures higher than 25°C.
- Risk of runaway.

### 3.2.2    Electropolishing (EP)

*Composition*
- ~9 Volumes of $H_2SO_4$ (buffer, very viscous).
- ~1 Volume of HF (complexant of $Nb^{5+}$, dissolves the oxide layer formed due to the high potential applied to $Nb^0$).
- Variation of composition allows us to adjust the etching rate.

*Pros* (in ideal conditions, i.e. viscous layer present)
- This is really 'polishing', not sensitive to crystallographic defects – it produces a smooth surface.
- Should not be sensitive to the cathode–anode distance – the same etching rate everywhere.
- It gives (but not always!) the best ever $E_{acc}^{max}$ ~ 45 MV·m$^{-1}$ (Tesla shape → ~180 mT).

*Cons*
- It is not possible to reach an ideal state in most of our processing conditions.
- Very sensitive to stirring condition, temperature, and aging of the mixture.
- Not very reproducible.
- Safety issues (acid mixture sensitive to water, $H_2$ evolution, etc.).

*Caution!*
- If $T$ increases: the etching rate increases but there is also a risk of pitting, H loading and HF evolution.
- If $V$ increases: the etching rate increases but there is also a risk of pitting, the generation of sulphur particles and sensitivity to the cathode–anode distance.

### 3.2.3    New tumbling developments

Tumbling, or Centrifugal Barrel Polishing (CBP), was initially an industrial process. It has been applied to cavities for decades at KEK or DESY, but was not optimized to reduce the damaged layer. Work on this topic has been resumed at Fermilab, with extra care being taken to use less aggressive polishing media, inspired by the slurry used in metallography. Although the process is not fully optimized, and still has a lot of intermediate steps, it has been possible to prepare mirror-like finish cavities with a very good RF performance after CPB and only 20 μm of EP. The removal of welding defects, including some pits, seems very effective, and improvements in performance have been observed (CBP + 20 μm EP showed improved performance compared to heavy EP) [35, 36].

Performing CBP saves a huge quantity of acid mixture and is preferable from the safety point of view. The evolution of the process to hydrogen-free processing, inspired by the work at Higuchi, is under way [37]. Nevertheless, this process is not applicable to every possible cavity shape. Studies should be conducted to see if the polishing step can be applied initially on the niobium sheets before forming.

Thorough control might become prohibitive when dealing with large batches, so once a vendor has been qualified, we can consider taking only random samples. The statistical efficiency of such sampling is well established in industry and can be relied on (see, e.g., Ref. [38]).

**3.3  Surface damage**

*3.3.1  'Skin-pass', the damaged layer, and recrystallization*

The surface state of the niobium sheets is directly linked to the industrial procedure: after lamination, the sheets are never completely flat. They have to be straightened by a superficial lamination process that is called 'skin-pass'. Skin-pass tends to concentrate damage (i.e. dislocations, displaced atoms, etc.) near the surface (see Figs 7 and 8).

Differential strain has very bad consequences during the recrystallization steps that will be performed afterwards: the material is strongly deformed at the surface but much less so in its core. In the case of weak deformations, we are close to critical hardening and thus we risk an exaggerated growth of the grains in the core, while the grains near the surface remain small.

Moreover, this situation is aggravated due to *texture*. Indeed, the texture at the core of well-crystallized niobium is rather like type (111), while the superficial texture because of the skin-pass is more like (100) and spreads in across about 200 μm. This latter orientation does not seem to recrystallize as well as the (111)[3] It will become part of the so-called 'damaged layer' of 100–200 μm, which has to be removed from the surface of the cavity by chemical abrasion. This also explains the patchy nature of the distribution of dislocations: some grains recover well during recrystallization, while others retain damage, depending on their initial orientation (see the discussion below on the influence of baking).

---

[3] T. Bieler (MSU), K.T. Hartwig (Texas AM), R. Crooks (Black Lab) ; personal communications. Note that recrystallization also occurs during the 800°C, 2 h treatments usually applied to get rid of hydrogen after heavy (electro-) chemical polishing.

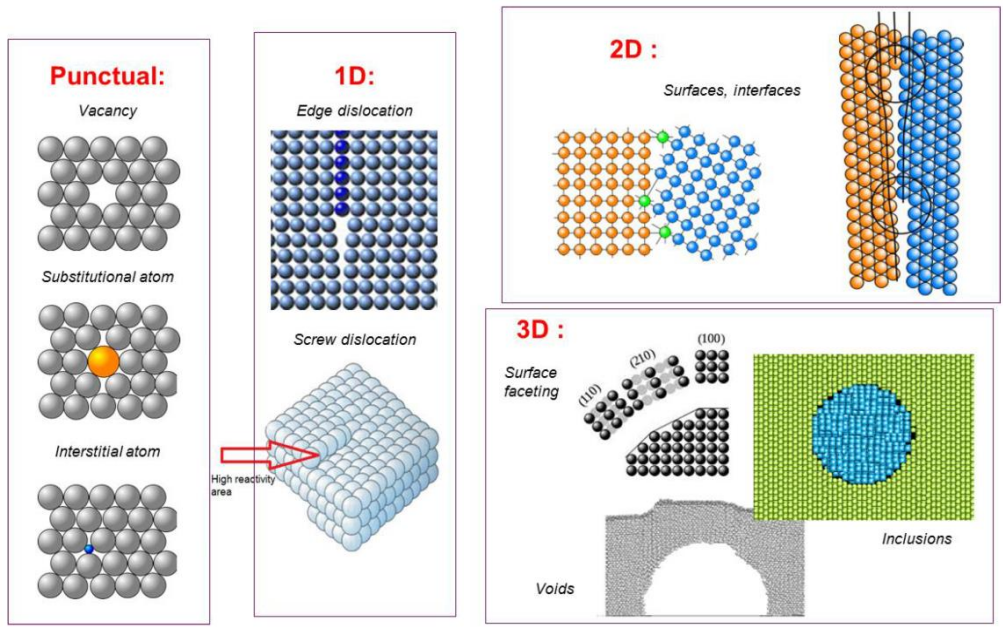

**Fig. 7:** Examples of crystalline defects frequently encountered in materials

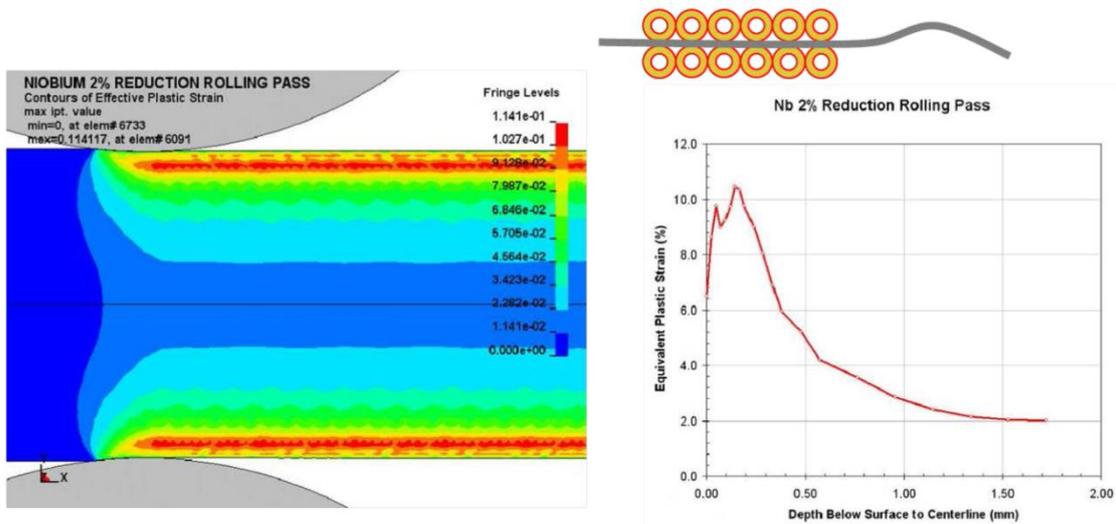

**Fig. 8:** A finite element simulation of the 2% reduction of a 3.5 mm sheet with 1 cm diameter rolls. The strain is concentrated in the near-surface region (red). The localized strain exceeds the average by a factor of 5. (Courtesy of Non-Linear Engineering, LLC and [39].)

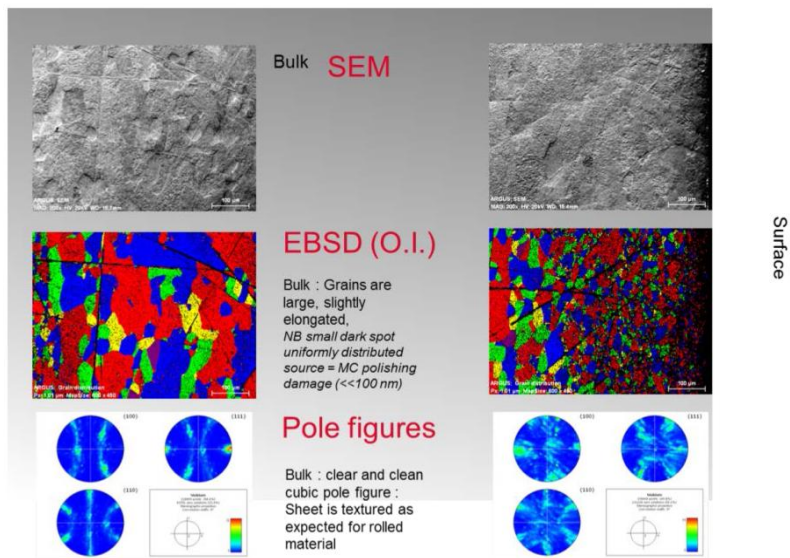

**Fig. 9:** A cross-section of a Nb sheet as received (after mechanical–chemical polishing of the cut surface). The top view is the SEM. The middle view shows the false colour map obtained from the electron backscattered diffraction diagram. Electrons penetrate ~100 nm of the surface and when the crystal is too disordered, no diffraction pattern can be interpreted, and only a black spot is shown. The lower part view is a pole figure. A perfect crystal would only exhibit dots. When the material is too disordered, no clear pole figure can be observed. On the left-hand side, one can observe the core of the sheet. The grains are large, with a uniform size. They are slightly elongated, which is often the case for a rolled sheet with incomplete recrystallization. The pole figure is also classical for a rolled sheet with a heavy texture. On the right-hand side, on can observe very small grains, many of them so disordered that no diffraction is observed. The pole figure, is blurred, indicating a highly disordered material. The first 100 μm of the surface are the most affected, but the damaged layer extends 200–300 μm deep under the surface.

The depth of the damage may also increase somewhat during deep drawing, although the deformation generally remains below 30% and is easily recovered. Additional thermal strain can also appear if the welding is not properly conducted. On the other hand, a series of uncontrolled chemical polishing operations ('pickling'/brightening by the manufacturer, treatments before welding, etc.) that can reduce it are applied during the fabrication process. Up to now, there has not been a systematic study aimed at minimizing this layer, although it might have important consequences for the surface treatments applied to cavities for large-scale production.

Up to the present day, manufacturers are still far from being able to control the production of Nb in a reproducible way, and in the context of large orders (ILC type), there is certainly reason to tighten our specifications on some realistic values and remain highly watchful as to the material that is being delivered.

### 3.3.2   *Damage, the baking effect, and superconductivity*

Until recently, it was very difficult to make a link between crystallographic observations and the superconducting properties of the cavities. Recent studies by A. Romanenko and collaborators have opened up this pathway.

Romanenko worked on samples cut out of cavities (small grain or large grain, BCP or EP) where temperature mapping allowed the selection of cold and hot spots, and he made a thorough comparison of hot versus cold spots from the structural, morphological, and chemical points of view [40-43].

It ensued that the main difference between cold and hot spots was observed by *Electron Backscatter Diffraction (EBSD)*, which is an electron diffraction technique that probes to more or less the same thickness as the field penetration depth. Indexation of the diffraction pattern gives the exact orientation of each grain, but one can also gain access to an evaluation of the local misorientation and dislocation density. Romanenko observed that hot spots tend to exhibit higher misorientation angles and a higher dislocation density.

Fig. shows the distribution of the misorientation angles for a picture formed on various types of samples, before and after baking. Except for small-grain BCP, misorientation shifts are reduced upon baking. Small-grain cavities indeed show no improved performance after baking.

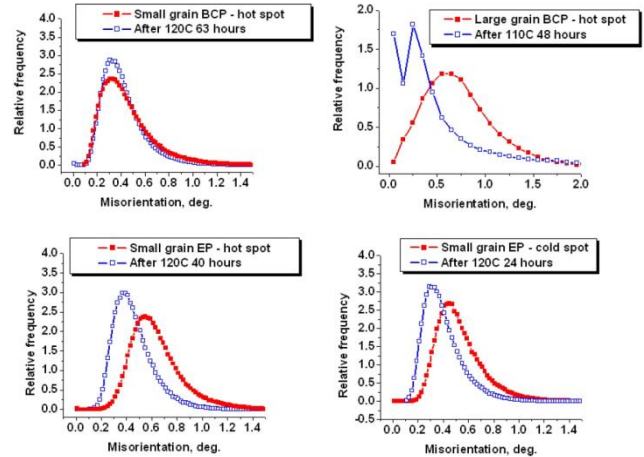

**Fig. 10:** Local misorientation shifts in cavity samples due to mild baking. Measurements of the misorientation distribution are performed on the same sample before and after baking. Except for small-grain BCP, misorientation shifts are reduced upon baking. Small-grain BCP cavities indeed show no improved performance after baking. (Courtesy of A. Romanenko, after [41].)

If the residual rolling strain is at the origin of the patchy repartition of hot spots, then this behaviour is very coherent with the nature of the damaged layer that we have described above: because of the surface texture, some grains with specific orientations resist recrystallization and retain a high density of dislocation.

Now we must consider how a high density of dislocations can interfere with superconductivity. Here again, Romanenko has explored a very interesting aspect. He found that Nb samples cut from cavities that have undergone cold RF tests exhibit the characteristic features of hydride precipitates, and that in accordance with the density of dislocation, the density of hydrides is higher on hot spots than it is on cold spots. Indeed, hydrogen is known to segregate along dislocations, forming 'Cottrell clouds'. This local high concentration of hydrogen promotes the formation of hydrides on cold RF testing. Nevertheless, these characteristic flakes are not systematically observed on H-rich niobium and their appearance probably depends a lot on the surface preparation and temperature cycles.

The presence of hydrides has also been demonstrated by Raman spectroscopy, where the characteristic vibrational bands from NbH and $NbH_2$ have been observed on cold as well as hot spots[44].

A. Grassellino has also observed the same samples in the presence of a magnetic field with muon spin rotation spectroscopy [43].

In this technique, positive, polarized muons, with their spin aligned to the beam direction, are implanted on the surface of the sample (typically 300 µm). In a metal, muons tend to position themselves in an interstitial site of the crystalline lattice. The frequency of precession of the muon's magnetic moment is of the order of a number of microseconds and depends on the local magnetic field. Muons decay in some hundreds of microseconds into a positron and two neutrinos. The positron is preferentially emitted in the direction of the muon spin at the moment of its disintegration. Taking into consideration the initial direction of the muon magnetic moment and the time interval between the moment of injection and the moment of muon decay, one can calculate how the precession frequency is influenced by the local magnetic field. Moreover, the decay signal also contains information about the magnetic volume fraction related to a particular frequency. Basically, the measurement is put in the form of a signal $A(t)$, where the frequency of the oscillation gives the amplitude of the local field, while its amplitude is a function of the magnetic volume fraction. This technique can be used to probe – among other things – the field penetration in superconductors.

Several 'hot' and 'cold' samples have been studied in the presence of a magnetic field perpendicular to the surface. In principle, this field geometry is different from the field repartition inside RF cavities, where the field is parallel to the surface, but we will see in Section 3.4 that the existence of a perpendicular field component is possible due to the surface morphology.

The onset of flux entry measured by A. Grassellino strongly correlates with the onset of RF losses as measured during the RF test by thermometry. **Erreur ! Source du renvoi introuvable.** shows a typical example for a sample cut out of a 'hot spot'. After baking, the onset of field penetration is shifted towards higher field for hot as well as cold spots, which is highly consistent with Romanenko's observations on dislocation density.

It is very difficult at this stage to evaluate the exact mechanism that affects superconductivity. Niobium hydrides are poor superconductors or normal conductors, depending on their exact stoichiometry. One can suppose that they will favour preferential field penetration. Nevertheless, hydrides, the density of dislocations and other types of defects are reduced by baking in a similar way, and the presence of hydrides might be a collateral effect rather than the origin of the losses.

This series of experimental observations shows indubitably that dislocations might play a paramount role in the superconducting behaviour of niobium, and that evaluation of their origin as the study of a possible cure to lower their density might be necessary to increase the reproducibility of cavity production.

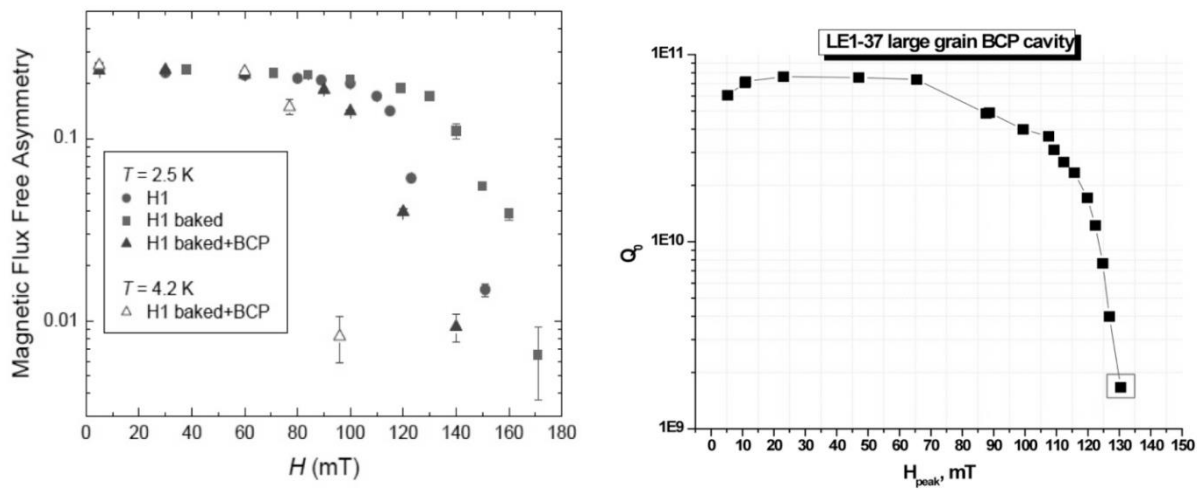

**Fig. 11:** Left: the amplitude of the asymmetry signal $A(t)$, which is proportional to the volume fraction of the sample not containing magnetic flux. When vortices start to enter the sample, this fraction decreases. In this example, one can clearly see that baking increases the onset of field penetration, while subsequent polishing restores the sample close to its initial behaviour. Right: the behaviour of the cavity from which the sample was cut out (after [43]).

### 3.3.3  *Tunnelling spectroscopy (point contact tunnelling)*

Point contact spectroscopy is a conductance measurement performed by bringing the tip from a tunnel microscope into contact with the surface. When the tip is just in contact with the surface, the measurement yields high resistance due to the insulating oxide layer. In this case, the interface layer between the superconductor and the oxide is also probed during the measurement. On the other hand, it is possible to produce a low-resistance measurement by pushing the tip across the oxide layer. In this case, the superconductor is probed somewhat more deeply and can be considered to be representative of the bulk material. By comparing the low- and high-resistance results, one can identify problems related to the actual surface of the superconductor.

The first measurements on monocrystals of niobium with the same RRR as the material of the cavities did not show any dramatic difference in the gap compared to pure metal: 1.55 meV for both low and high impedance. The low-impedance (i.e. bulk) measurements show a purely BCS behaviour.

In the high-impedance measurement, however (with the oxide), there is an increase in the density of states of the quasi-particles (smearing) that cannot be explained as being merely an effect of the measurements. Moreover, the signal is enlarged compared to a purely BCS behaviour, which means that the mechanism involved is not the proximity effect but comes from a breaking of the Cooper pairs [45].

An additional correction needs to be introduced. At this stage, the only mechanism that adequately reflects these results is the theory of Shiba [46] that refers to the inelastic scattering of quasi-particles on magnetic impurities in the case of strong coupling: the curve is adjusted by means of a so-called 'pair-breaking parameter', $\Gamma$, that is related to the concentration of the diffusion centres.

After baking, the parameter $\Gamma$ decreases sharply. It is also very interesting that there is no difference at the interface between vacuum baking and baking in air, just as is observed on cavities.

In addition, recent statistical analysis on the point contact tunnelling spectrum measured on chemically polished cavity-grade Nb samples reveals that the pair-breaking parameter, $\Gamma$, depends on the resistance of the junction: low-resistance junctions ($R \leq 100$ $\Omega$) exhibit a bulk Nb superconducting gap of 1.55 meV and almost no pair breaking, whereas higher-resistance junctions ($R \geq 1$ k$\Omega$) show higher values of $\Gamma$ (see fig. 12). A preliminary interpretation of these results is that magnetic impurities are localized in the uppermost defective $Nb_2O_5$ oxide and/or at the interface with the underlying sub-oxides forms ($NbO_2$, $NbO$).

The origin of the baking effect described earlier might also be found in a local reorganization at the interface, inducing a lower concentration of localized magnetic moments or a weaker coupling with the underlying Nb superconductor. Because of the low dimensionality of the interface and the highly defective nature of the $Nb_2O_5$, diffusion can still make a difference even at low temperature.

The same approach has been applied to coupons cut out of cavities (cold and hot spots).

In the case of hot spots, the density of state is so high at the Fermi level that it forms a central peak in the conductance spectra (around zero bias). This peak remains even if the superconductivity is killed by the presence of an external magnetic field, and can be attributed to the Kondo effect, due to the presence of localized magnetic moments [47, 48]. This feature appears for ~30–50% of the junctions measured on hot spots, as compared to ~5% on samples of cold spots.

The existence of these magnetic impurities has been now confirmed by SQUID magnetometry. The interesting fact is that the density of the magnetic impurities depends strongly on the initial surface preparation of the samples and the integrated magnetic signal intensity scales with the surface area of the sample, which confirms that they are localized in a very superficial surface layer [47, 48]. We should consider these results in the light of the magnetic measurements described in [49], that also brought to the fore the existence of magnetic impurities in the first 10 nm of the niobium surface.

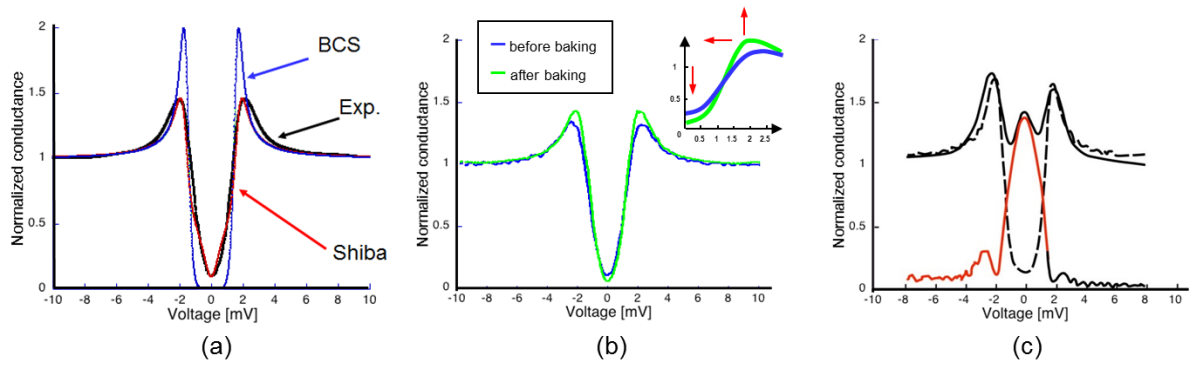

**Fig. 12:** Various examples of high-impedance conductance curves measured at 1.6 K. All curves show a superconducting gap Δ equal to 1.55 meV, as expected for pure Nb. (a) In black, conductance curves measured at 1.65 K at the surface of a monocrystalline niobium sample that was electropolished before baking; in red, the fit of the experimental data with the Shiba model – the adjustment is done using a parameter Γ that takes into account the inelastic scattering of quasi-particles by magnetic impurities. In blue, for the sake of comparison, the expected BCS signal for the same superconducting gap Δ. (b) A comparison of the same sample before (in blue) and after baking (in green). The insert shows an enlargement of the 0–2.5 mV part. After baking, the scattering is reduced and the shape of the spectrum becomes more BCS-like (red arrows). (c) A typical conductance curve observed on hot spots (in black). Deconvolution shows that it is probably the superposition of a localized state at zero field to the superconducting gap.

It is indeed known that niobium oxides are sub-stoichiometric and have a variable density of oxygen vacancies. During baking, radical changes occur in the metal/oxide interface: $Nb_2O_5$ dissociates into $NbO_2$ and $NbO$, as we were able to establish by other means [50]. One can imagine that the density of magnetic centres decreases considerably during this transformation. However, in comparison to earlier work [51], the magnetic signal intensity is too important to be caused just by sub-stoichiometric $Nb_2O_{(5-x)}$ $4d^1$ electrons (roughly by a factor of between 10 and 50, depending on the surface treatment), in qualitative agreement with earlier work done by Casalbuoni [49]. Up to now, the additional contribution remains a mystery. It might be possible that O–H–Nb vacancy complexes present close to the surface ( see Section 4.2.4 from [52]) can exhibit some magnetic behaviour, but this still needs to be explored.

The existence of magnetic impurities results in a non-zero density of state at the Fermi level (the so-called 'gapless superconductivity regime'), which is liable to produce dissipations and suppress the superheating field [48, 53, 54]. In particular, in Ref. [54], a simple model is considered with uniform spatial distribution of magnetic and non-magnetic impurities (i.e. within ξ from the surface), which has the advantage of being valid at arbitrary frequency, κ (the Ginzburg–Landau parameter, GL), temperature and scattering rate compared to Δ. The authors computed the surface impedance ($R_S$) in the presence of magnetic impurities in the Shiba approximation. This reveals a saturation of $R_S$ at low temperature, suggesting that magnetic impurities can be responsible for an appreciable fraction of the residual resistance.

One might wonder how these results correlate with the observation of early field penetration in high dislocation density areas, as discussed in Section 3.3.2. One can only speculate, but areas with a high dislocation density are liable to oxidize faster than the remaining surface. For instance, the emergence of a screw dislocation on the surface provides a small group of isolated atoms ( see Fig. 13), which is highly reactive and would provide a perfect nucleation site for oxide growth [55]. Pitting and etching also occur preferentially in high dislocation density areas [56-58]. Moreover, angle-resolved XPS studies show that the thickness of the oxide layer is not uniform [19]. All these 'ingredients' are compatible with oxide becoming thicker or more defective in the emerging dislocation areas.

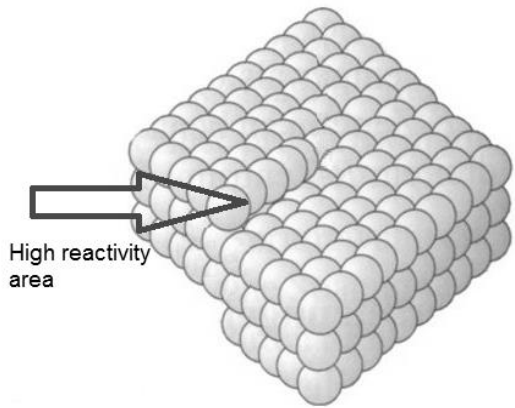

**Fig. 13:** A schematic of the emergence point of a screw dislocation

We believe that here lies an interesting direction for further research, not only from a practical point of view (process optimization), but also for a deeper fundamental understanding of this aspect of superconductivity.

### 3.4 Surface morphology and quench

Even when the dissipation phenomena at high field are poorly understood, it is relatively easy to establish the *quench*: this is the generalized transition of the superconductor to the normal state. The cavity then finds itself detuned and most of the incident power is reflected. Because of the high BCS resistance, at high frequency the generalized heating is often at the origin of the quench. But at 1.3 GHz the thermal instability can in general be attributed to the presence of a localized defect (of the size of some tens of micrometres in diameter) that may provoke a localized increase in the magnetic field and/or in temperature and thus lead to a quench. Chasing defects the size of some microns on surfaces the size of a square metre turns out to be quite a challenge: there are a great many possible defects, but only some effectively influence the functioning of the cavity. Why and how? This is the both practical and fundamental issue at stake in this chapter.

#### 3.4.1 Surface morphology

The topography of the surface can also be invoked to explain the quench, as it might locally increase the magnetic field; for example, on the steps that appear at the grain boundaries (see Fig.). If the local field exceeds $H_C$, an area of the material may transit to normal state and start to dissipate strongly. This phenomenon was originally proposed at Cornell to explain the thermal dissipation at high field ($Q$-slope) [59], which was invalidated later by the studies on baking [60].

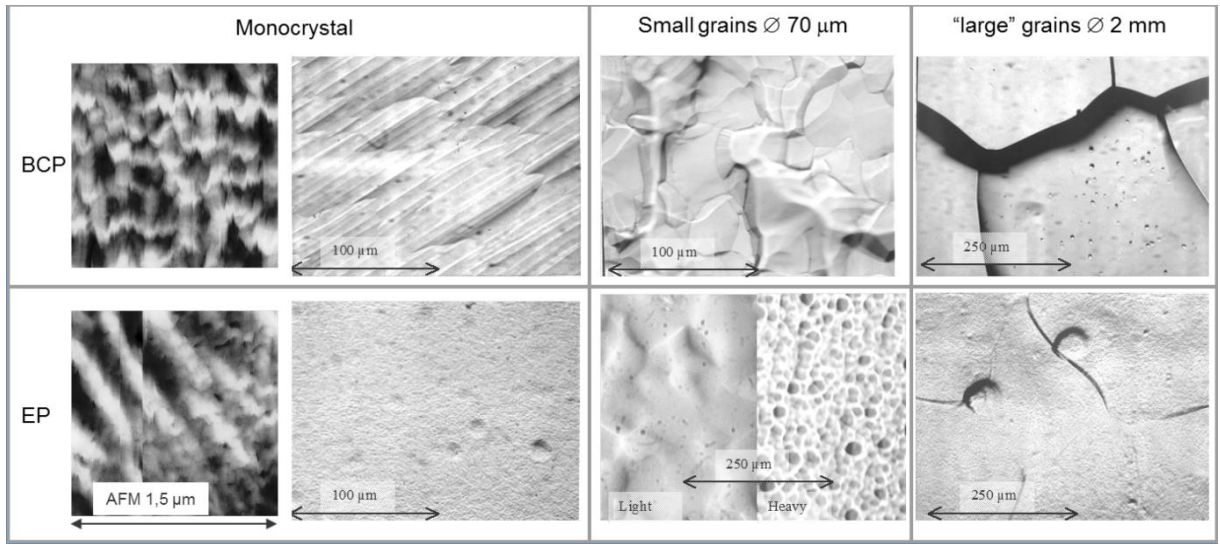

**Fig. 14:** The morphology brought about by the different surface treatments greatly depends on the initial crystalline state. Chemical 'polishing' is not a true polishing: it reveals crystallographic defects such as grain boundaries or the emergence of dislocation lines. Not all the grains etch at the same speed. Therefore steps appear that are about 20% of the size of the grains. At a nanometric scale, though, it is not possible to distinguish the effect of one treatment from the other.

Still, this approach remains of interest to explain the quench, provided that the local increase of the field can be evaluated correctly. The thermal calculations show that the quench is due to a localized defect with sides no larger than some microns, and therefore very difficult to localize on a macroscopic object such as a cavity.

We therefore decided to attempt a morphological characterization of the surface, to see whether there might be a correlation between the quench and a particular kind of surface defect. There were two problems, as follows.

- How do we measure the surface near a real quench, given that we cannot measure the RF behaviour of small samples directly?[4] We need to develop an *in situ* and non-destructive measuring technique.

- Which morphological parameter should we consider? As we can see in Fig., the conventional roughness parameters are not well adapted to our problem; because very different profiles for the electromagnetic properties may have the same roughness values (arithmetic $R_a$ or quadratic $R_q$). Nevertheless, profile (b) should lead to a far larger increase of the field on the peaks than profile (a).

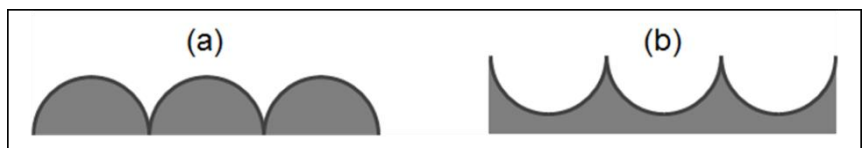

**Fig. 15:** Two examples of surface morphology with the same roughness parameters, but very different behaviour with respect to the EM field

---

[4] Many attempts have been made to design cavities that can withstand samples. In fact, most of these cavities are at higher frequencies and do not give a large enough precision to extrapolate to the behaviour of niobium at 1.3 GHz.

*3.4.1.1 Temperature maps*

Several temperature-mapping systems exist. Here, we give the example of a superfluid helium temperature-measuring arm that was developed some years ago. It allows detection of the position of the quench *in situ* by means of thermal diffusion through the cavity wall (see Fig.). The distance between the sensors, of the order of a centimetre, and the angular resolution of the motor give a precision of 2–3 degrees at the level of the equator. Once the location of the quench has been determined, we may study the inner surface. Optical methods such as endoscopy are not well adapted: at low magnification it is not possible to distinguish a detail of a size of ~10 μm; at high magnification, there is not enough depth of field. We therefore need to apply other methods (a mechanical sensor (profilometer), electron microscopy, etc.) without having to cut a part of the cavity! There are several methods for making replicas, but in our case it was necessary to ensure fidelity better than a micron, on uneven surfaces. During the RF tests of the cavities, we made a temperature map in order to localize the quench, followed by making a replica at the site of the quench and at two reference surfaces (far from the quench), to see if we could find a correlation between the quench and a certain kind of defect or a specific morphology.

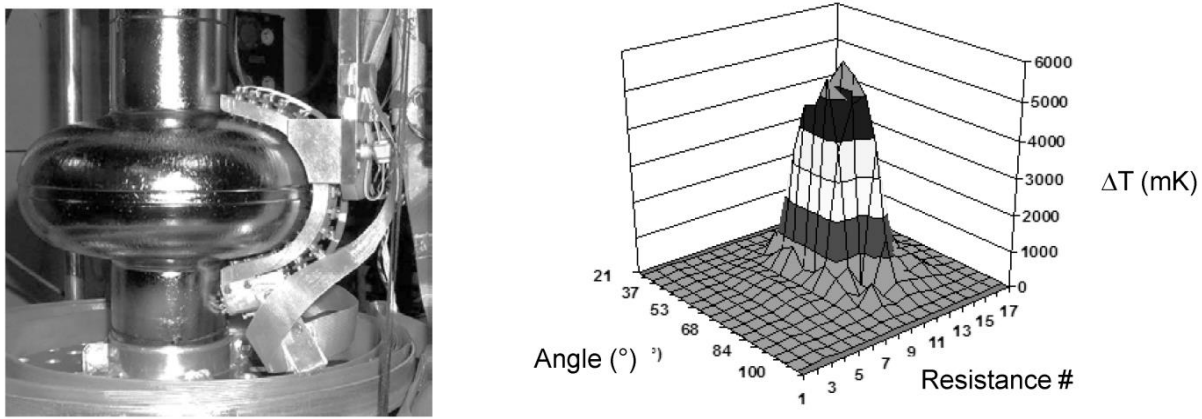

**Fig. 16:** The mobile temperature-measuring arm and the map obtained during an RF test of a cavity at 2 K. To obtain this sort of map, one needs to place oneself at a field value just before the quench.

As we saw before, conventional roughness parameters are not suitable. We need to measure a quantity that takes into account the curvature radius of the observed topography, and the increase of the field is caused by sharp features on the surface.

The measured curvature radius depends on the scale of observation.[5] To start with, we have studied the phenomenon at a scale of one tenth of a micron, for which we can use profilometry. We then could extend the study if a higher resolution proved necessary.

*3.4.1.2 Topological analysis*

There exist topological methods that allow us to make a global estimate of the surface state that is more accurate than classical roughness.[6] Among these, there is analysis by 'equivalent conformal elliptical structure' [62]. In this technique, the surface relief is discomposed into small pyramids and the surface of each face is projected on to a plane perpendicular to the three directions of the space (with the same origin). In each direction, the projections tend to exhibit elliptical shapes, which allow a 3D ellipsoid to be reconstructed. The three axes of the obtained ellipsoids are topologically related to the surface morphology and it is clear that the 'steep' steps will tend to increase the vertical axis of the ellipsoid. This technique provides a tool that helps us to characterize the tendency of the surface to exhibit sharp features.

---

[5] The dependence of the data on the scale of measurement is a widespread problem, but is often ignored with respect to roughness. An approach using fractals allows the scale to be overcome: one may, for example, express quadratic roughness as $x^n$, where $x$ is the scale of observation and $n$ is a fractal number. This approach has been successfully applied to lightly de-polished silicon wafers [50], but so far we have not succeeded in applying it to polycrystalline niobium, probably due to the great variability of the morphology of each grain.

[6] The work of Tian *et al.* [61] H. Tian, et al., "A novel approach to characterizing the surface topography of niobium superconducting radio frequency (SRF) accelerator cavities". Applied Surface Science, 2010., based on power spectral density, also allows us to characterize the surface morphology at different scales.

Moreover, the demagnetization factor $D$, which is the inverse of the (magnetic) field enhancement factor $\beta$, is analytically calculable for an ellipsoid (see Section 3.4).

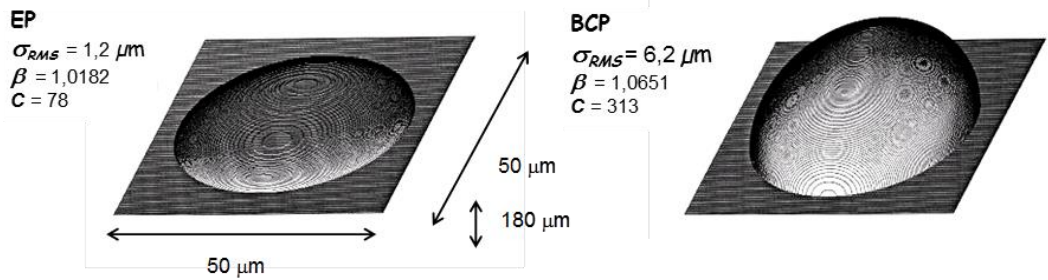

**Fig. 17:** Equivalent ellipsoids deduced from measurement of electropolished (EP) and chemically polished (BCP) surfaces, respectively. First, the surface is measured using a profilometer on a $50 \times 50$ µm$^2$ surface with 9 µm steps. Then the topological analysis is performed on the acquisition file. We have also indicated the quadratic roughness ($\sigma$), the average field enhancement factor ($\beta$) and the value of the half-axis $C$. The analysed surface contains ~1000 grains.

In the example shown in Fig., we analysed un-annealed material (grain Ø ~ 70 µm) with steps of 9 µm. If one measures the roughness at a much smaller scale, and/or on material with larger grains, the differences are far less pronounced and the measurements on etched niobium are close to that of electropolished metal. This clearly shows that the roughness is mainly due to differences in the etching rate of the grains and to the resulting steps in the case of BCP.

The values of the increase of the average field (some per cent) are not enough to explain a quench, but they do give a qualitative assessment of the roughness: obviously BCP produces 'sharper' reliefs than EP. This comprehensive approach allows for a general qualification of a surface treatment. The same approach can be used to qualify an individual step. This has been done by INFN-Legnaro to assess the influence of distribution of steps on the field; each step is modelled by a half-ellipsoid[7]. They showed that only a single step, with a high form factor and placed perpendicularly with respect to the field, is needed to trigger a quench.

We arrived at the same conclusions using 2D modelling of the field with FemLab™, but also by means of direct measurements on a step (see below).

### 3.4.2    *Replicas at the quench site*

To study the morphological evolution of the quench site during successive treatments, a method of replica-making proved to be the most practical. We opted for a technique validated at the University of Besançon, which guaranteed us the desired level of reliability. This is a very inexpensive method that has already proven itself before (on Formula 1 engines!). Following a short period of collaborative work to verify the validity of the method on niobium samples, we adapted the technique to cavities, by developing a tracking system suited to the convex and closed shape of the cavity. This technique has also been used more recently to study etching pits found in the heat-affected zone of the welding seam [63].

---

[7] V. Palmieri, C. Roncolato, personal communication

Fig.8 illustrates the various stages of this technique: having spotted the quench on the outside of the cavity, it is positioned in such a way that the meridian passing through this point is in a low position. We introduce a little calibrated ball that allows us to identify the meridian and that gives us a size scale. Then, with the aid of a mirror, we take a photograph of the inner surface. There is a slight error due to the fact that we project a curved surface on to the plane of the photograph, but the degree of accuracy is sufficient for our needs. Next, we place a soft polymer (of the 'dental impressions' type) to make a negative replica of the surface. We then transpose this replica positively on a hard polymer, which is analysed using a profilometer. The area we want to study is identified with the help of photographs and the placement of coordinates thereupon. As one can see on the measurements, the existence of a grain that is prominent compared to the others is evident, although because of the lighting it is very difficult to distinguish it in the photographs.

This technique offers a significant advantage: it is very inexpensive and relatively easy to implement. It has allowed us to open up a new direction in the exploration of the inner surface of cavities that until then had been neglected.

From the topography of the surface, we introduced the real profile of the steps in a 2D model that allows us to reproduce the local increase of the field due to the morphology. This model can still be improved, because it does not take into account the finite size of the grain in the direction perpendicular to the field, but it allows us to assess the overall influence of the morphology.

### 3.4.3    *Modelling the generated field*

The models are 2D and refer to a step placed on an infinite plane. The magnetic field is parallel to the surface of the plane and perpendicular to the step. The calculations show that the smaller the radius of curvature, the larger is the increase of the field [64, 65]. But measurements taken directly on cuts of the replica show that the radius stays large (~50 μm), a value that we have retained in the subsequent calculations. It seems that the next most important parameters are the slope and the height of the edge of the grain. Fig.9 shows the different results that were obtained: one enters a profile determined from the experimental measurement (a). One then determines the average increase of the field with respect to the mean field established far from the disturbance (b). One then can calculate the contribution of the step to the dissipated power (blue curve in (c)) compared to the power dissipated across the basal plane (green curve).

It is clear that the influence of the step arises suddenly when approaching the transition field. This behaviour is quite similar to that observed in the cavities, where a set of hot spots can be seen to appear on the surface associated with the *Q*-slope. But at the moment of the quench, a single point ( not necessarily the hot spots already shown previously) becomes much hotter and causes the generalized transition. In the case of the grain that is studied in Figure 18 (for a sheet with a thickness of 2.8 mm) a thermal calculation shows that one can stabilize two normal areas on the 'noses' of the steps (an area with a diameter of ~1 μm, where $T < T_C$, but $H > H_C$) up to ~142 mW. With but 1 mW more power, the material quenches completely.

Fig.20 shows the same type of calculation (2D), but now coupling the RF behaviour to the thermal behaviour of Nb. These calculations show that if the dissipation is correctly evacuated to the helium bath, it is possible to maintain an superconducting state up to a certain field, but that a slight increase in the field (a few tenths of a millitesla) induces thermal runaway.

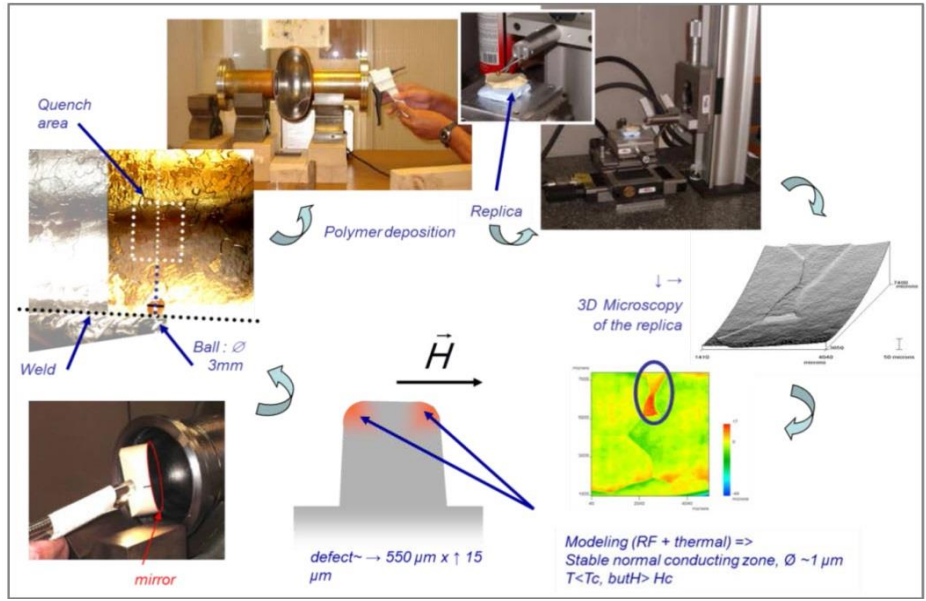

**Fig. 18:** The different stages in the characterization of the internal surface of cavities. Because of the artificial enlargement of the vertical axis, the 'prominent' grains are easier to highlight. At normal scale, a step of the size of 10 μm on a grain that is several hundreds of micrometres in size is quite difficult to distinguish.

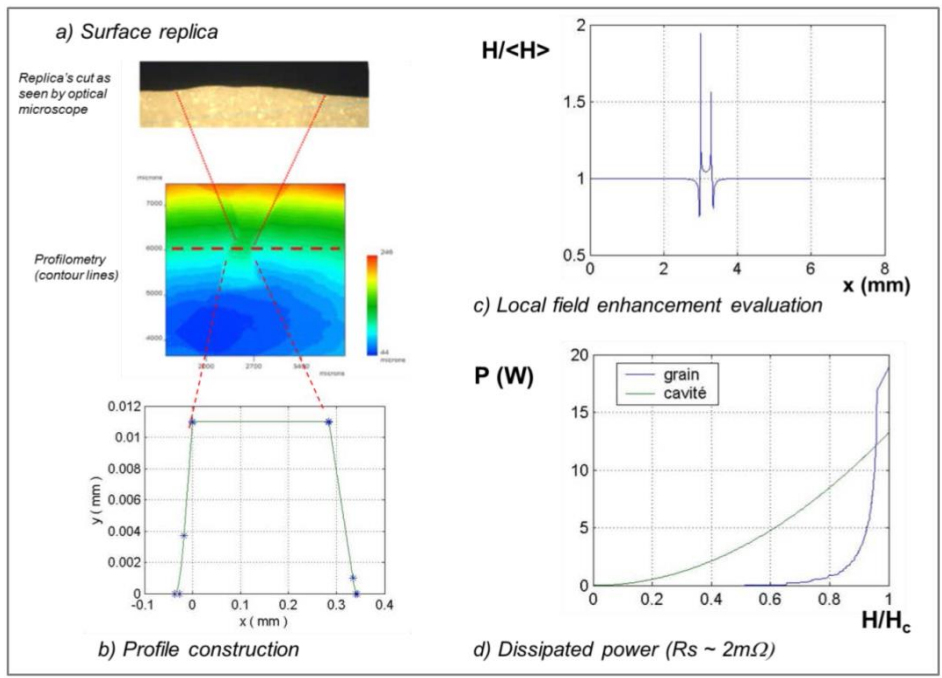

**Fig. 19:** A calculation of the effect of the surface morphology on the local increase of the magnetic field and the power dissipation (see text). (a) A micrograph of a cutting of the replica along the dotted line on the contour (measured before cutting). (b) The profile used for calculations. (c) Calculations of the local increase in the field induced by the defect. (d) The power dissipation of the cavity free of defects (in green) and of the default only (in blue). We see that the default becomes dominant at high field and can induce a quench.

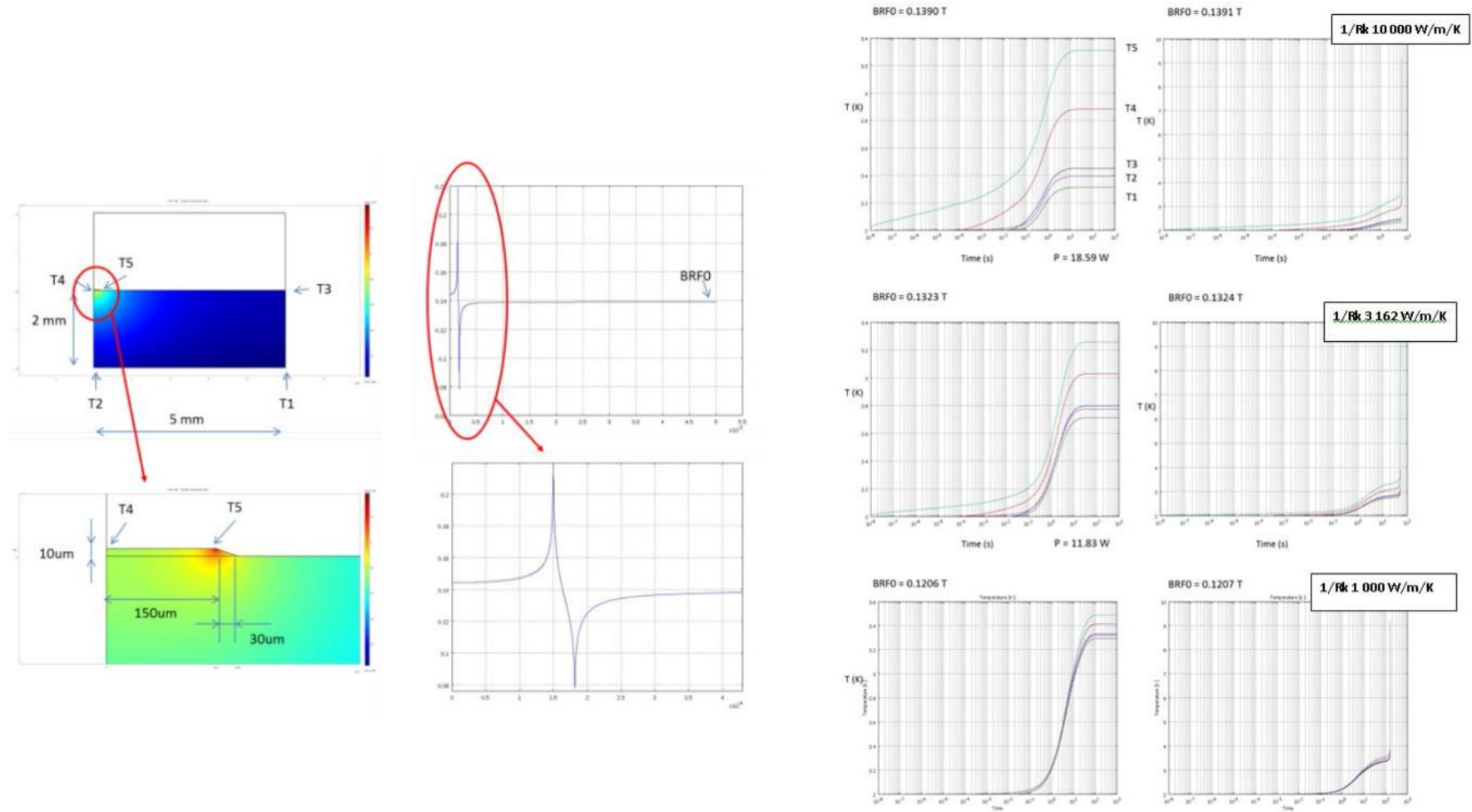

**Fig. 20:** The amount of power that can be transferred to the bath is clearly limited by the interface transfer for a given thermal conductivity. An increase in the purity of the Nb allows us to reduce the interstitial content, which acts as scattering centres for (thermal) conduction electrons. It increases the thermal conductivity and allows us to transfer the dissipated power better, and hence to obtain a higher field for an 'equivalent' defect. The quench happens on the edge of the defect because of both morphological field enhancement and temperature enhancement: thermomagnetic quench at $H < H_C$ and $T < T_C$.

Fig. shows the evolution of this quench after a chemical treatment of 20 μm. The temperature maps show that the site of the quench has moved by several centimetres. However, this chemistry is not sufficient to fundamentally alter the topography of the grain under consideration; hence the factor with which the field increases. But the new site clearly shows a new step, for which the field increase factor is higher than at the former quench location. Strictly speaking, we cannot prove that the step did not already exist beforehand. It is, however, easy to show that it is always the step with the highest $\beta$ that will quench first.

The surface morphology thus explains some premature quenches observed on cavities treated by BCP, where enhanced roughness appears close to the welding seam, due to recrystallization of large grains with different orientations.

The influence of grain boundaries and the morphology on field penetration have also been studied by magneto-optics [66] and critical current measurements.

The study of bi-grains shows that there is indeed a preferential penetration of the field at the grain boundaries, but only when the field is perfectly aligned with the boundary plane, and for (static) fields much higher than the peak fields obtained in RF; that is, under conditions that are quite far from the configuration of the cavities. In combination with the results on the 'monocrystal' cavities, this shows that, although grain boundaries are an area of weakened superconductivity, they do not dominate the dissipative phenomena in SRF [66].

An experiment on a monocrystal in which a groove was deliberately dug, however, shows that when the field is applied parallel to the surface but perpendicular to the groove, a significant vertical field component appears on the edges of the groove (see Fig.). This proves that our 2D model is realistic. To try to quantify the influence of a grain with finite and realistic dimensions, we also need 3D modelling [66].

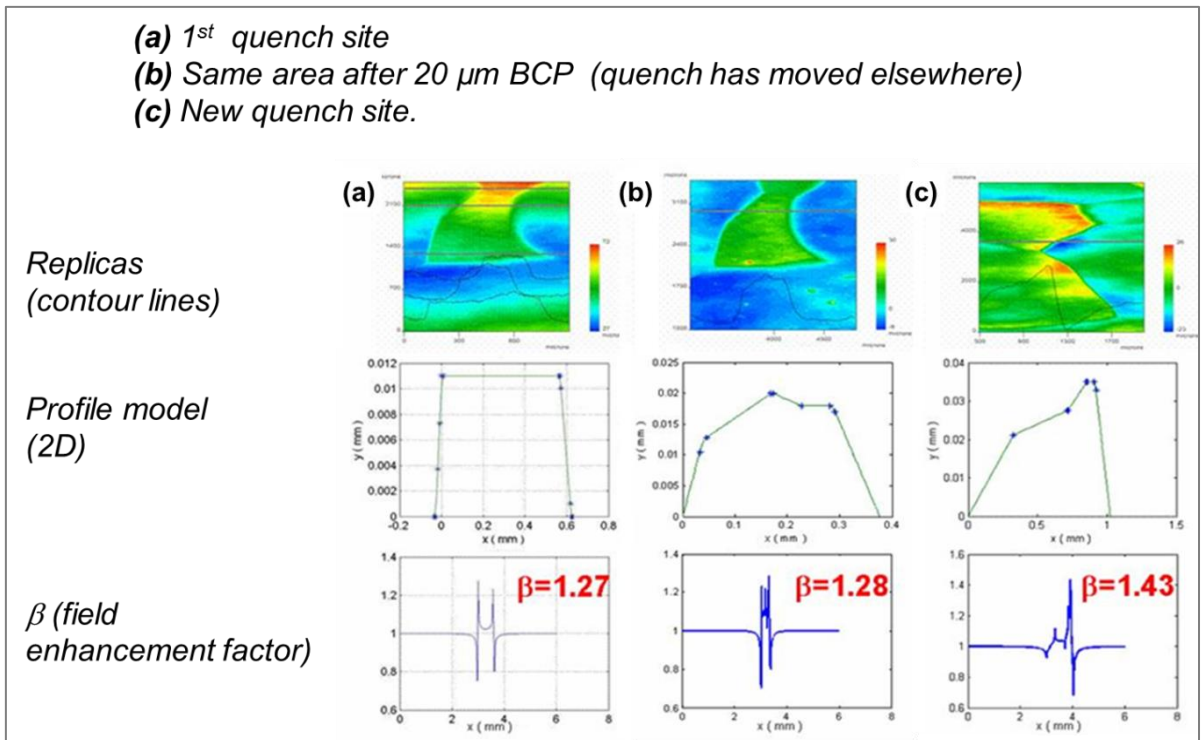

**Fig. 21:** (a) The first site of a quench ($E_Q$ = 24.72 MV·m$^{-1}$ at 1.7 K). (b) The same zone after 20 μm of BCP (the quench has moved elsewhere). (c) The location of the new quench ($E_Q$ = 24.94 MV·m$^{-1}$ at 1.7 K).

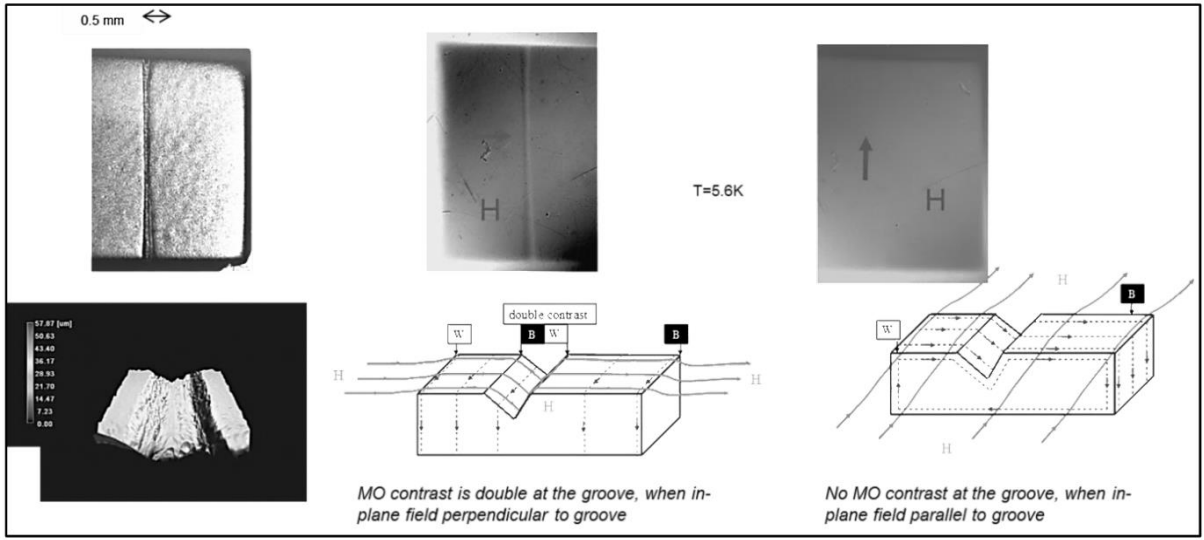

**Fig. 22:** The effect of an artificial groove on the field distribution on the surface of a monocrystalline sample (after [66]). Left: optical microscopy and 3D profilometry. Right, top: images obtained by magneto-optical (MO) contrast for field parallel (centre) or perpendicular (right) to the groove. Bottom : schematic repartition of the field at the origin of the MO contrast.

We remark that the results obtained on monocrystalline cavities are very recent and quite unexpected. Indeed, in numerous superconducting applications, the grain boundaries exhibit weakened, or even an absence of, superconductivity, and have a predominant influence on the properties of the superconductor. We have therefore tried to explore this aspect in some detail.

### 3.4.4 Welding and roughness

The grain size greatly affects the roughness. Near the weld, the grains of the heat-affected zone have diameters close to 0.5–1.0 cm. The heights of the steps, as well as the parameters of the equivalent ellipsoids, increase correspondingly. The local increase of the field, of the order of 40–50%, starts to become significant.

Using the 'ellipsoid' approach described in Section 3.4.3, we were nevertheless able to estimate the demagnetization coefficient of a single step or a surface. Indeed, the computation of the demagnetization factor for an ellipsoid is relatively easy:

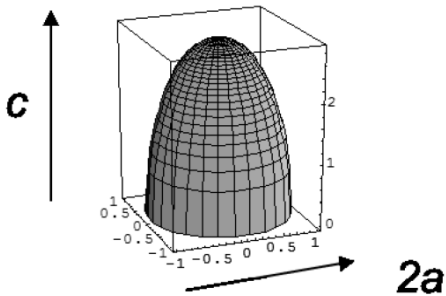

$$D_a = \frac{1}{1-m^2}\left\{1 - \frac{m}{\sqrt{1-m^2}}\arccos m\right\}, \qquad (4)$$

where $m = c/a$.

Table 1 shows the evaluation of the demagnetization factor estimated after the ellipsoid parameters as directly measured on surfaces with various roughness characteristics. Note that from the topological point of view, a hole is equivalent to a protrusion, so the same approach can be used to study etching pits.

**Table 1:** Roughness parameter and demagnetization *C* factors for ellipsoids measured on standard surfaces

| Parameter | Etching (BCP) | | | Electropolishing (EP) | |
|---|---|---|---|---|---|
| | Small-grained material | Annealed, far from the weld | Thermally affected zone (near the weld) | Mean value | Weld defect* $C \sim 50$ μm $2A \sim 200$ μm |
| Φ grains | 70 μm | 1–2 mm | 0.5–1.0 cm | 1 mm → 1 cm | - |
| $R_a$ | 1–2 μm | 4–8 μm | 40–80 μm | ~1 μm | - |
| $C$ | ~300 | ~90–100 | ~350 | ~70 | 50 |
| $\beta = 1/D$ | 1065 | 1028 | **1.4** | 1018 | **1.9!*** |

*A defect associated with a hot spot on a cavity quenching at 15 MV·m$^{-1}$, observed at Fermilab [63, 67] (cf. Fig.).

These simple calculations can also be confirmed using the finite element approach. Fig. shows the modelling of an ellipsoid of similar dimensions and shows similar modelling for a sharp-edged fracture. Here again, we can observe that the most important part is the area with the sharpest curvature.

Similar 3D simulations have also been conducted at Cornell [68] and show that field enhancement factors between 1.5 and 2 can easily be reached in defects that are very small compared to the cavity size.

In the case of an electric field, enhancement by a factor of 50–100 is often necessary to model field emission, and it is doubtful that surface morphology plays an important role; whereas a magnetic field and enhancement by a factor of 2 will have serious consequences for the transition field.

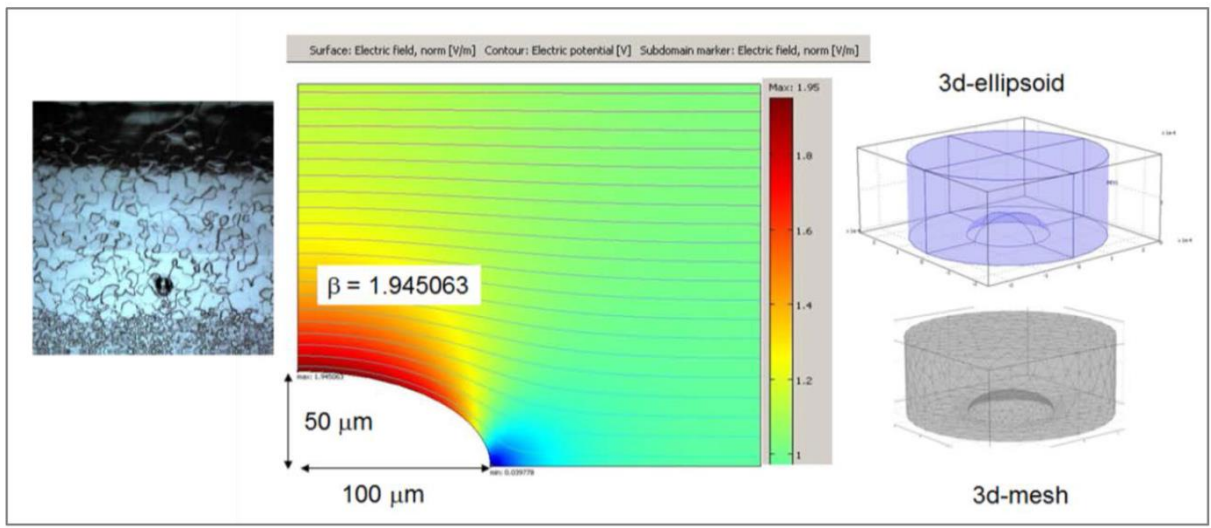

**Fig. 23:** The finite element simulation of the field enhancement factor for an ellipsoid of 50 μm × 100 μm (courtesy of Z. Insepov).

Note that the surface morphology does not matter much for field emission, whereas it can be the dominant effect for the appearance of a quench.

Cavities treated by electropolishing, followed by a moderate baking, show quench thresholds that are systematically higher than those of cavities treated in a conventional way, by chemical polishing and baking (35–40 MV·m$^{-1}$ and 25–30 MV·m$^{-1}$, respectively). Only a few individual cavities (in thousands of tests) have succeeded in reaching ~40 MV·m$^{-1}$ after chemical polishing, but it is not possible to check the reproducibility of these results. Among these were mainly seamless cavities or cavities without boundaries (prepared by hydroforming or spinning; monocrystalline cavities) [69, 70].

*It is therefore our belief that all the techniques that maintain welds and grain boundaries are likely to give rise to a premature quench in case of BCP treatment.* 'Large-grain' cavities do not come with any particular advantage in the case of BCP surface treatment, except in terms of the cost of the niobium supply. Electropolishing does not make all the steps on the surface disappear, but as the radius of curvature on the steps is much larger, the field enhancement factor is not so high. That could explain the relatively better results obtained with this surface treatment.

### 3.5 Conclusion on surface defects

When the accelerating gradient is below 40 MV·m$^{-1}$ (~170 mT), one can be sure of the presence of a local defect, wherever it is an inclusion, a pit or relief, or a local variation of composition. If the cavity quenches at less than 15–20 MV·m$^{-1}$, then this defect is 50–100 μm sized [30] and can be seen with the naked eye (seeTable **1**Table 2).

The origin of the inclusion is seldom the initial Nb sheet due to its very high purity; but since it is a very soft material, the dust particles commonly found in a workshop environment will tend to become embedded. Special care has to be taken with regard to the cleanliness and smoothness of the tools that may come in contact with Nb, a habit that is not yet common in many industrial plants and that deserves close attention.

Pits can originate from by means of different mechanisms.

(1) *During solidification of a melted area*. Light elements (C, O, N, etc.) tend to diffuse towards the melted part during cool-down, and thus get concentrated in the last liquid parts. If the vacuum environment during melting is not good, then it can reach the point at which the light elements form bubbles inside the material. The typical feature of such a pit is sharp edges (due to faceting during crystallization), and high concentrations of C and O in the walls of the pit ([71] and personal unpublished results). The critical steps are the melting of the ingot and the welding.

(2) *Pit corrosion during etching*. If strain remains in local spots, it will favour pit corrosion. The typical features of an etching pit are that it is ~100 μm in depth, 50 μm in diameter, smooth, and very shiny [72]. These local strain areas form in particular with thermal stress, such as can appear after welding if the cool-down conditions are too abrupt.

One can see that welding is a critical part of the cavity fabrication process that calls for particular care.

**Table 2:** Examples of typical large defects

| Defect type | origin | Quench field |
|---|---|---|
| Bubble, seen by XR ∅ 0.5 mm | Saclay, Bad EB welding | ~ 12 MV/m |
| Bad vacuum during EB welding | Experience at Saclay and DESY | 20-25 MV/m |
| 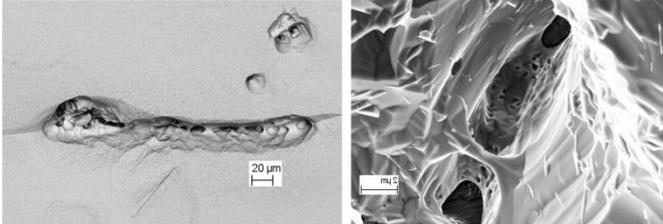 | Desy Bad vaccum EB welding ? | ~ 16 MV/m |
| 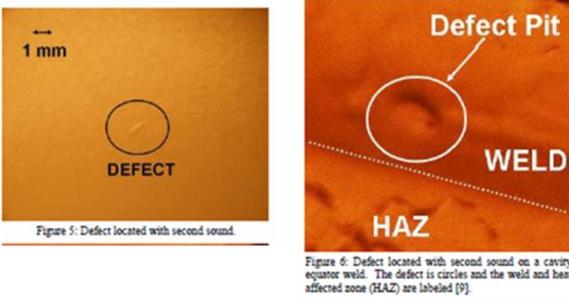 | DESY Bump, defect in the deep drawing die | ~ 20MV/m |
| Ta inclusion (un-cleaned rolling machine) | DESY | 8 to 14 MV/m |
| 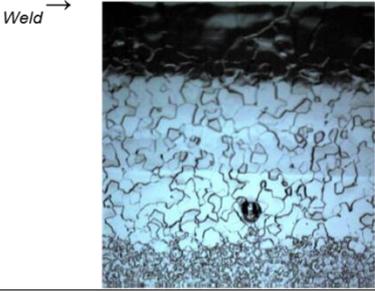 | FNAL Pit, in the HAZ | ~ 15 MV/m |

### 3.6  Niobium: from bulk to thin films

The deposition of thin films on to copper is not a recent idea. Indeed, from the superconducting point of view, only a few hundreds nanometres of superconducting material is necessary. Copper presents a good mechanical support with fine thermal conductivity. Such a technology would decrease the cavity fabrication costs considerably. Indeed, several accelerators worldwide are based on magnetron-sputtered Nb films (e.g. LHC and Soleil). This technique is well adapted for circular machines where high accelerating gradients are unnecessary, but until recently, although a large amount of R&D has been conducted, thin film technology has not been applicable to high-gradient applications (see fig. 24). This is probably related to the fact that sputtered films have a highly defective crystalline structure: a lot of displaced/foreign atoms, very small grains, high dislocation density, and so on.

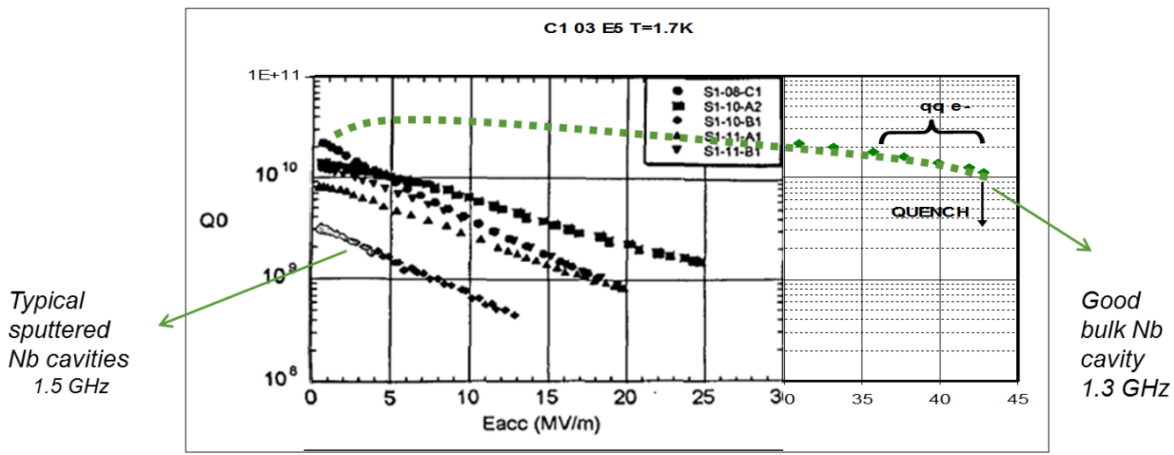

**Fig. 24:** A comparison between typical sputtered Nb films deposited on copper and bulk Nb cavities

Only recently, progress in deposition techniques has allowed the deposition of 'bulk-like' films that might result in a breakthrough in the fabrication of high-performance Nb cavities (see, e.g., Refs. [73-75]).

Nevertheless, bulk as well as bulk-like niobium are close to the ultimate limits of this superconductor (transition to the normal state). Only a different type of superconductor can overcome this limitation (see below).

# 4   After (bulk) niobium

After nearly 30 years, despite several attempts at using other superconductors, bulk niobium is still unmatched for high-gradient applications. Nevertheless, recent results seem to indicate that we are about to reach the material's ultimate limits, notably with several observations, since 2008, of the location of quenches randomly fluctuating for a given field. This shows that we are no longer limited by a point defect leading to a premature transition, but that several areas of the surface have the same transition probability. This field is found around $43 \pm 2$ MV·m$^{-1}$ (1830 ± 85 Oe, 'Tesla shape', RRR 300[8]), a weaker field than predicted by the earliest theories.

For a long time, the failure of these attempts was thought to be related to difficulties in preparing the superconductors themselves. Indeed, the superconductors with the highest values of $T_C$ are generally compounds with several possible phases and the superconducting phase often corresponds to a fairly narrow region of concentration. As soon as we have a region with a lot of defects (e.g. grain boundaries), the stoichiometry and the regularity of the lattice are no longer guaranteed, which may result in a degraded superconductivity.

## 4.1   Ultimate limits: recalls

In principle, the theoretical maximum of the accelerating field is obtained when the magnetic component of the electromagnetic field at the surface of the cavity reaches the superconductor's transition field at the operating temperature. The precise real limit is not known: the exact mechanisms of radiofrequency dissipation at the operating temperature (2–4 K) are not well understood, and most of the established models are valid only close to $T_C$.

Niobium is a classical *type II* superconductor, which means that it is well described by the BCS theory (Bardeen–Cooper–Schrieffer), and its extension GLAG (Ginzburg, Landau, Abrikosov,

---
[8] The record accelerating field has been reached at Cornell on a low-losses cavity made of Nb RRR 500. With that peculiar shape, the exceptional 59 MV·m$^{-1}$ corresponds to 2065 Oe.

Gor'kov), which describes the type II behaviour. The specific description of a superconductor in a.c. (RF) has been proposed by Gorter and Casimir (1934) with the two-fluids model: charge carriers are divided into two subsystems, the superconducting carriers (Cooper pairs) of density $n_s$ and the normal electrons of density $n$.

A type II superconductor can exhibit three states. At low temperature, it is in the Meissner state. In the presence of an external magnetic field, a screening current appears at the surface of the superconductor (at its penetration depth $\lambda$) and produces a magnetic moment opposite to the external field up to its first critical field, $H_{C1}$, where $H_{C1}$ corresponds to the transition from the purely superconducting state to a mixed state, where both normal and superconducting areas coexist. The normal areas consist of field lines called 'vortices' ('vortex' in the singular), surrounded by screening currents. The second critical field, $H_{C2}$, corresponds to the transition from the mixed state to the normal state. Other transition fields that can be considered are $H_C$, the critical thermodynamic field, and $H_{SH}$, the superheating field (see

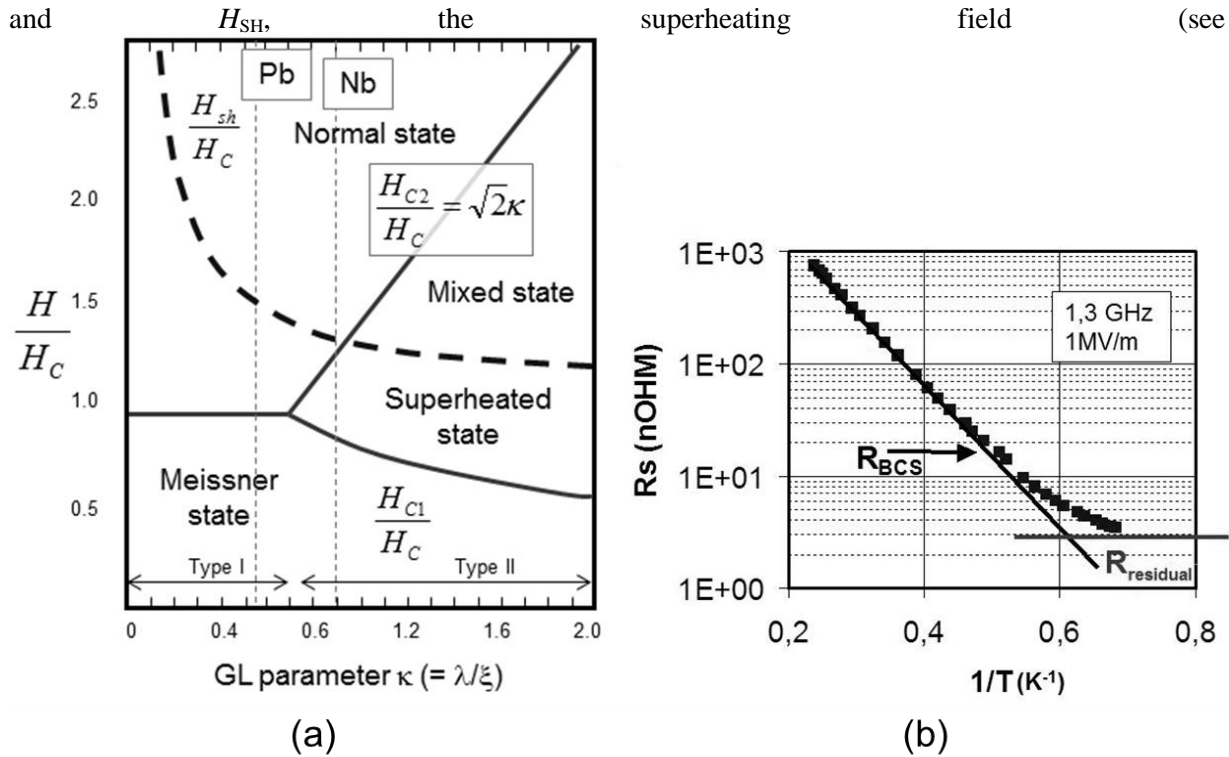

(a)                (b)

Fig.).

The question of which transition field ($H_C$, $H_{C1}$, $H_{C2}$, or $H_{SH}$) should be considered is still controversial and will be discussed in this section. For other details on the general and mathematical description of superconducting states, the reader is invited to consult textbooks in superconductivity (see also G. Ciovati's tutorial in this volume).

### 4.1.1 *Superheating field*

It is possible to observe a metastable state where superconductivity persists at higher fields than $H_C$ (type I superconductors) or $H_{C1}$ (type II), up to the so-called 'superheating' field $H_{SH}$. In the 1960s, the field $H_{SH}$ was estimated on the basis of thermodynamic considerations related to the surface energy

[2]. It follows the dotted curve in

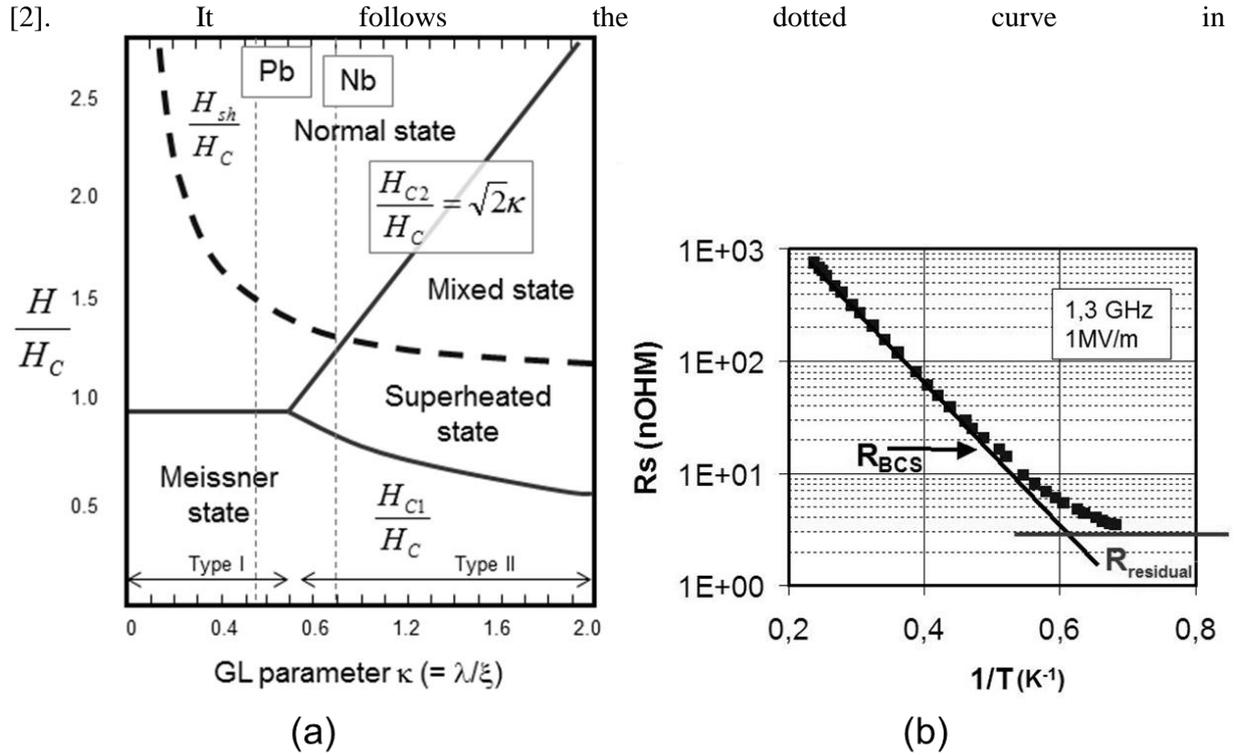

(a)                      (b)

Fig.(a). It was measured near $T_C$ on several type I and Type II superconductors in pulsed resonators for various frequencies [76]. It is easier to observe this effect in RF than in d.c., because apparently surface defects play a lesser role in nucleating the RF magnetic transition than they do in nucleating the d.c. transition [76].

A superconductor can be described by its Ginzburg–Landau parameter, which is the ratio of the field penetration depth to the coherence length of the Cooper pairs: $\kappa = \lambda/\xi$.

According to the values of $\kappa$, the 'superheating' field can be approximated by the following expressions [2]:

$$H_{SH} \approx \frac{0.89}{\sqrt{\kappa_{GL}}} H_C \text{ if } \kappa \ll 1, \tag{5}$$

$$H_{SH} \approx 1.2 H_C \text{ if } \kappa \sim 1, \tag{6}$$

$$H_{SH} \approx 0.75 H_C \text{ if } \kappa \gg 1, \tag{7}$$

where $H_C$ is the critical thermodynamic field.

From the superheating model, one should expect $E_{acc} > 50$–$60$ MV·m$^{-1}$ for niobium (with $Q = 10^{11}$, at 2 K and 1.3 GHz).

For several decades, the commonly accepted explanation was that the field reverses every $10^{-9}$ s, whereas it takes $10^{-6}$ s to reach the nucleation of a normal zone. Therefore there is not enough time to see it nucleate. But even in his paper [76], Yogi claims that vortex nucleation dominates at lower temperatures and that individual vortex nucleation takes less than $10^{-9}$ s. Except very close to $T_C$, he always observed the RF transition at a lower field intensity than the ideal $H_{SH}$ value, and the temperature dependence indicated a line nucleation model; that is, vortex nucleation.

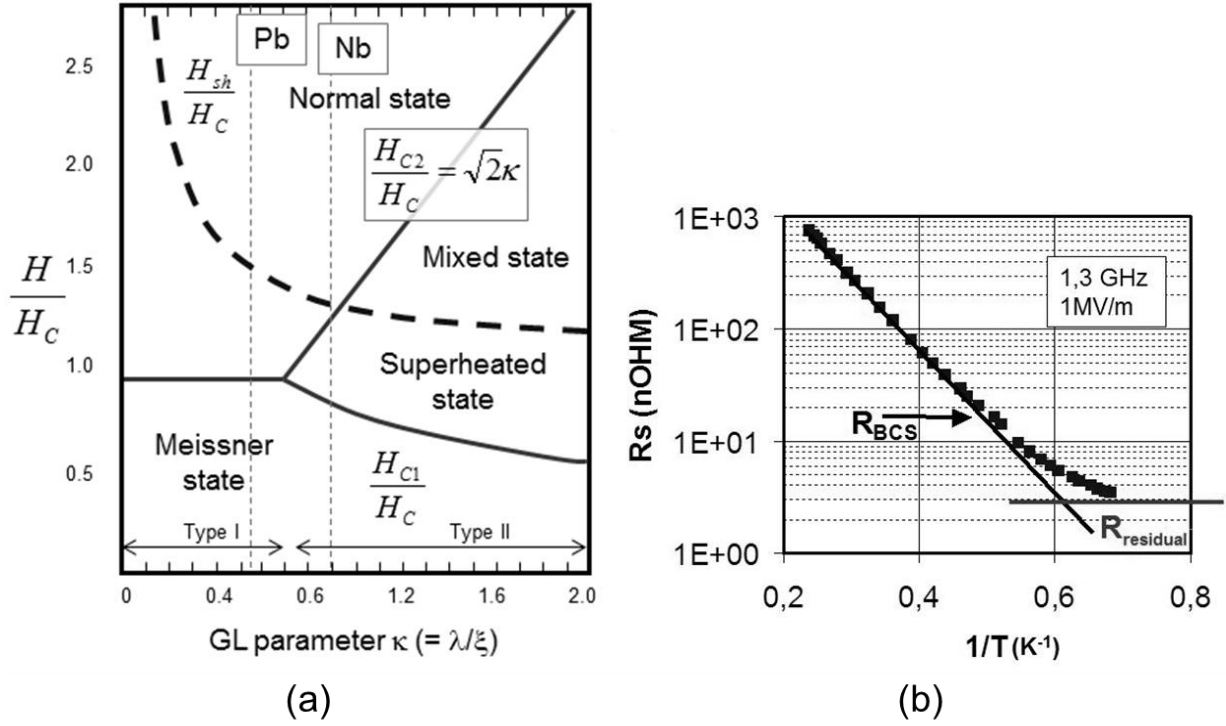

**Fig. 25:** The peculiarities of RF superconductivity. (a) In RF (dotted curve), one observes superconducting behaviour at higher fields than in d.c. situations (continuous curves). The 'superheating' field $H_{SH}$ is defined following thermodynamic arguments (see Ref. [2]), but obviously other less-defined mechanisms occur concurrently. (b) The resistance of a RF superconductor is not zero. At low field, the behaviour of the surface resistance follows that predicted by the BCS theory up to about 2 K, after which it gets dominated by the residual resistance $R_{res}$. At higher field, there is no valid model yet.

Note that $H_{SH}$ does not depend on $H_{C1}$ or on $H_{C2}$. Materials that are good superconductors for applications at direct current (e.g. magnetic coils) are most often 'bad' in RF. Indeed, the pinning centres of vortices (*dislocations*, precipitates, etc.) that allow one to obtain high values of $H_{C2}$ are very dissipative defects in RF.

*4.1.2   Vortex nucleation*

Nevertheless, the superheating model is based on the premise that the superconductor is 'defect free', a state difficult to attain in the 'real' world. Recently, the applicability of this model has been questioned. If defects are present at the surface, the penetration of individual vortices can be considerably faster (it is now estimated at ~$10^{-13}$ s [77-79]).

Penetration of vortices has now been proposed to explain the appearance of the '*Q*-slope' [79, 80] – that is, high field dissipation (see below) – and the limitation of the performance of cavities [81]. Vortices have to overcome a surface barrier (resulting from the conjunction of Meissner currents and surface image vortices), which only disappears at $H = H_C$. The surface barrier is reduced by the presence of defects [78].

When defects are present, then the Meissner state is expected to disappear at $H_{C1}^{RF}$ [9]. Note that $H_{C1}^{RF}$ is expected to be slightly higher than $H_{C1}^{DC}$ [82]. The ultimate field in this case would be directly linked to the thermodynamic field $H_C$, which is always difficult to measure precisely for type II superconductors.

---
[9] A. Gurevich, personal communication.

The origin of the presence of some vortices between $H_{C1}^{DC}$ and $H_C$ is not clear. Perhaps we are dealing here with residual magnetic field lines[10] that remained trapped during the cooling of the cavity.

The origin of the ultimate limitations in RF is far from being settled, especially at high fields and low temperature, where some approximations of the general theory are no longer valid [83]. The theoretical study of this subject has only just begun, partly because it is only nowadays that the cavities have intrinsic properties that are good enough to be able to test hypotheses dating back more than 50 years. However, recent propositions by A. Gurevich seem to point towards early magnetic field penetration issues and maybe a superheating field is not the right criterion for practical applications.

*4.1.3    Criteria for choosing a 'good' RF superconductor*

*A priori*, there are two partially contradictory aspects that come into play – the surface resistance (in order to have the highest possible quality coefficient $Q_0$) and the 'superheating' field $H_{SH}$, which affects the maximally possible accelerating field:

$$R_{BCS} = A(\lambda_L^4, \xi_F, \ell, \sqrt{\rho_n}) \frac{\omega^2}{T} e^{-\Delta/kT} . \qquad (8)$$

We see that in order to decrease $R_{BCS}$, we need to attain a maximum gap $\Delta$ (which requires a high $T_C$, because $T_C \sim 1.87\Delta/k_B$), a minimal penetration depth $\lambda_L$ and good electric conductivity at normal $\rho_n$. A low $R_{BCS}$ is a necessary but not a sufficient condition to get a low surface resistance. We must also ensure that we have a very weak residual resistance. For example, high-$T_C$ ceramic superconductors, with their weak links (e.g. grain boundaries in normal state), have a residual resistance at RF that is ~100 times higher than the resistance of copper. Therefore they are of no interest for RF applications, despite their high $T_C$ [84].

We are therefore striving to develop metal compounds with high values of $T_C$. Unfortunately, known high-$T_C$ compounds often also have a very high $\lambda_L$ (see Table 3). Getting a high-gradient accelerator automatically results from a compromise.

Table 3: The properties of several superconductors

| Material | $T_C$ (K) | $\rho_n$ (μΩ·cm) | $H_C$ (T)* | $H_{C1}$ (T)* | $H_{C2}$ (T)* | $\lambda_L$ (nm)* | Type |
|---|---|---|---|---|---|---|---|
| Pb | 7.1 | | 0.08 | n.a. | n.a. | 48 | I |

---

[10] Cryostats are magnetically shielded to protect them from the geomagnetic field.

| | | | | | | | |
|---|---|---|---|---|---|---|---|
| **Nb** | **9.22** | **2** | **0.2** | **0.17** | **0.4** | **40** | **II** |
| NbN | 17.1 | 70 | 0.23 | 0.02 | 15 | 200 | II |
| NbTiN | 16.5 | 35 | | 0.03 | | 151 | II |
| $Nb_3Sn$ | 18.3 | 20 | 0.54 | 0.05 | 30 | 85 | II |
| $V_3Si$ | 17 | | | | | | II |
| $Mo_3Re$ | 15 | | 0.43 | 0.03 | 3.5 | 140 | II |
| $Mg_2B_2$ | 40 | | 0.43 | 0.03 | 3.5 | 140 | II – two gaps |
| YBCO | 93 | | 1.4 | 0.01 | 100 | 150 | d-wave |

*At 0 K.

Compounds such as NbN or $Nb_3Sn$ do seem very attractive, but the attempts that have been made to prepare them, either by thermal means (from bulk niobium cavities) or by sputtering, have not yielded the expected results: they remain far behind bulk niobium. (For a review of these trials, see Ref. [85] and 26).

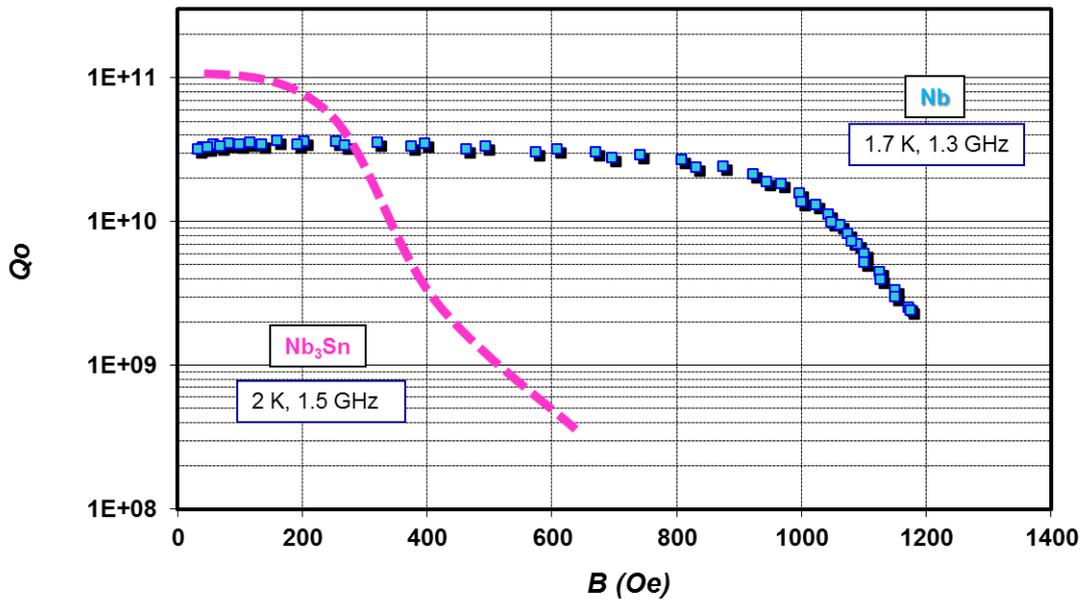

**Fig. 26:** A comparison between the maximum magnetic field for a $Nb_3Sn$ cavity obtained by $Sn^{vap}$ diffusion in Wuppertal, and a bulk Nb cavity tested at Saclay. The appearance of a slope on the $Nb_3Sn$ cavity around 30–35 mT may be related to the $H_{C1}$ value of $Nb_3Sn$ 50 mT at 0 K, while the slope for Nb around 100–120 mT might be related to its $H_{C1}$ value (170 mT at 0 K). The $Nb_3Sn$ curve is determined after Ref. [7], with geometrical ratios $H_p/E_{acc}$ = 4.7 mT for 1.5 GHz and $H_p/E_{acc}$ = 4.26 mT for 1.3 GHz.

Now if we consider that the penetration of vortices is promoted by surface defects, a superheating field is not liable to be reached in realistic conditions. Even on bulk niobium, where much progress has been done on surface preparation, we have several hints that crystalline defects as common as dislocations can promote field penetration (see Section 2.6). Even if the density of the defects can be reduced by appropriate surface preparation, not all of them can be removed. For instance, the density of dislocations in a well-recrystallized metal is still higher than $10^4$ cm·cm$^{-3}$ at room temperature (while it can reach $10^{12}$ cm·cm$^{-3}$ for heavily deformed metal).

The other direction that has been explored consists of depositing a thin film (a couple of microns) on a copper cavity (by sputtering, plasma, etc.), which is much cheaper and more favourable from a thermal point of view. These techniques are well suited to flat surfaces. In the cavity, layers generally come with a columnar structure, with porosity, defects, very small grains (100 nm), high residual stress, and a $H_{C1}$ value that is lower than that of the equivalent bulk material. It has been shown on niobium films (~1–5 µm) that the porosity increases with the roughness of the substrate, and this seems to cause an increase of the dissipation with the field (see, e.g., Ref. [86] and references therein). In this configuration, the material is entirely in a mixed state. It has been thought for a long time that these vortices were well anchored on the material defects and thus hardly dissipated. But as we have already mentioned before, the recent work of A. Gurevich has shown that the BCS resistance at high field is highly non-linear, and of magnetic origin [78, 80]. Moreover, grain size seems to influence the frequency at which vortices can be untrapped by an RF field [87]. This might explain why the technology of films on copper never produced very high field gradients.

This work perhaps also provides a good explanation why the $Nb_3Sn$ cavities did not lead to the expected results, as we can see in Fig.: at low field, the $Q_0$ value is very good, which means that the residual resistance and the BCS resistance are weak and that we did obtain the desired phase. But after no more than just a couple of $MV \cdot m^{-1}$, the $Q_0$ collapses and becomes far worse than that of the Nb.

Note that recent progress in deposition techniques seems to open the way for niobium thin films with bulk-like properties [88]. It will certainly lower the fabrication cost, but we will still be limited by the performance of niobium.

### 4.1.4 *The importance of $H_{C1}$*

So far, only the superheating field had been considered for the choice of superconductor, because it was assumed that the Meissner effect would be maintained above $H_{C1}$. But in realistic conditions, many surface defects still exist and can ease early penetration of vortices. If the actual limitation in cavities is related to the penetration of vortices via surface defects, and thus to the $H_{C1}$ of the material, it is not surprising that only niobium performs well: niobium has the highest value among all known superconductors!

It therefore seems unlikely that we will improve the performance of cavities using conventional superconductors. Only specific structures such as those proposed by Gurevich [89] are likely to go beyond the performance of niobium.

### 4.2 Superconducting nano-composites: an innovative path for the future of SRF

The recent work of A. Gurevich on the SRF has opened up a whole new approach, that might allow us to finally smash the monopoly of bulk niobium [89].

It involves the improvement of the performance of bulk niobium by 'shielding' it, using very thin films. Here, the order of magnitude is no longer the micron, but the nanometre. We will show that it will not only be possible to obtain higher-gradient accelerators, but also to diminish the cryogenic losses, depositing 30–50 nm of a high-$T_C$ superconductor such as $Nb_3Sn$ or $MgB_2$. An insulating layer of some 5–15 nm is required to decouple the two superconductors (in the sense of a Josephson junction). The presence of bulk niobium is still necessary, as the multilayers are too thin to fully screen the magnetic field.

The increase of the transition field originates from a property of very thin films: in this case, the field around a potential vortex decreases as $d/\pi$ instead of $\lambda$ (multiple interactions with the anti-vortex images), where $d$ is the film thickness and $\lambda$ is the penetration depth of the field. The lower critical field is much higher:

$$H_{C1} = \frac{\Phi_0}{4\pi\lambda}\left(\ln\frac{\lambda}{\xi} + 0.5\right) \tag{9}$$

becomes

$$H'_{C1} = \frac{2\Phi_0}{\pi d^2}\left(\ln\frac{d}{\xi} - 0.07\right), \tag{10}$$

where $\Phi_0$ is the magnetic flux quantum and $\xi$ is the Cooper pair coherence length. The surface barrier still exists, but it is also higher:

$$H_S \approx H_C = \frac{\Phi_0}{2\sqrt{2}\pi\lambda\xi} \tag{11}$$

becomes

$$H'_S \approx H'_C = \frac{\Phi_0}{2\pi d\xi}. \tag{12}$$

As an example:

- for NbN, where $\xi = 5$ nm, $H_{C1} = 0.02$ T, and $H_C = 0.23$ T, a layer of 20 nm will give $H'_{C1} = 4.2$ T and $H_S = 6.37$ T;
- for $N_3Sn$, where $\xi = 3$ nm, $H_{C1} = 0.05$ T, and $H_C = 0.54$ T, a layer of 50 nm will give $H'_{C1} = 1.4$ T;
- recall that for Nb, $\xi = 38$ nm, $H_{C1} = 0.17$ T, and $H_C = 0.18$ T [78, 89].

The surface resistance is also expected to diminish with the use of superconductors with high $T_C$. High(er)-$T_C$ superconductors have a BCS resistance that is much weaker than that of niobium. The experimental issue is then to find conditions of fabrication that are gentle enough to not introduce too many defects: extremely uniform layers, clean interfaces, few grain boundaries, little residual stress, and not too many foreign or displaced atoms, so that the residual resistance stays low. Fig. and 28 show recent experimental measurements on samples deposited using multilayers [90-92]. This topic now needs to be systematically developed, first to determine the optimum number/thickness distribution for the multilayer and, secondly, to adapt the deposition techniques to the building of such layers inside cavities.

## 5   Conclusion

Progress in SRF technology over the past 20 years has led us to the ultimate limits for niobium technology, laboratory-wise. A lot of work is still needed to be able to produce, at high yield, a large number of accelerator cavities with the same performance. Mastering the surface state – in terms of cleanliness, purity and crystalline perfection – seems to be of paramount importance. New technologies are on the threshold of supplanted bulk niobium. In the short term, new thin-film technologies will allow bulk-like Nb material to be deposited, rendering the SRF technology much cheaper, while in the longer term, multilayer structures, applicable to the upgrading of existing Nb structures, as well to new ones, will hopefully allow us to go towards higher field with reduced cryogenic losses.

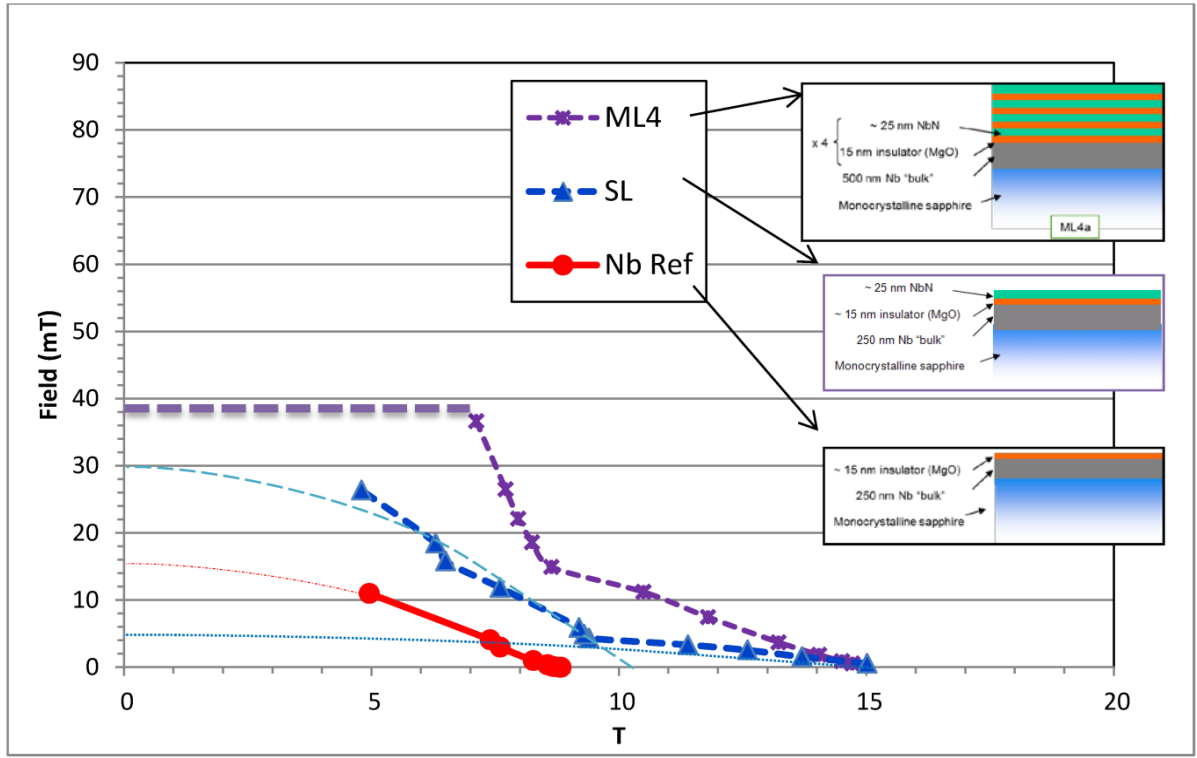

**Fig. 27:** The magnetic behaviour ($H_{C1}$) of various single and multilayer samples with 25 nm layers of NbN. The field at which vortices enter the sample is systematically higher than the Nb-only reference sample, demonstrating the screening effect of the layers.

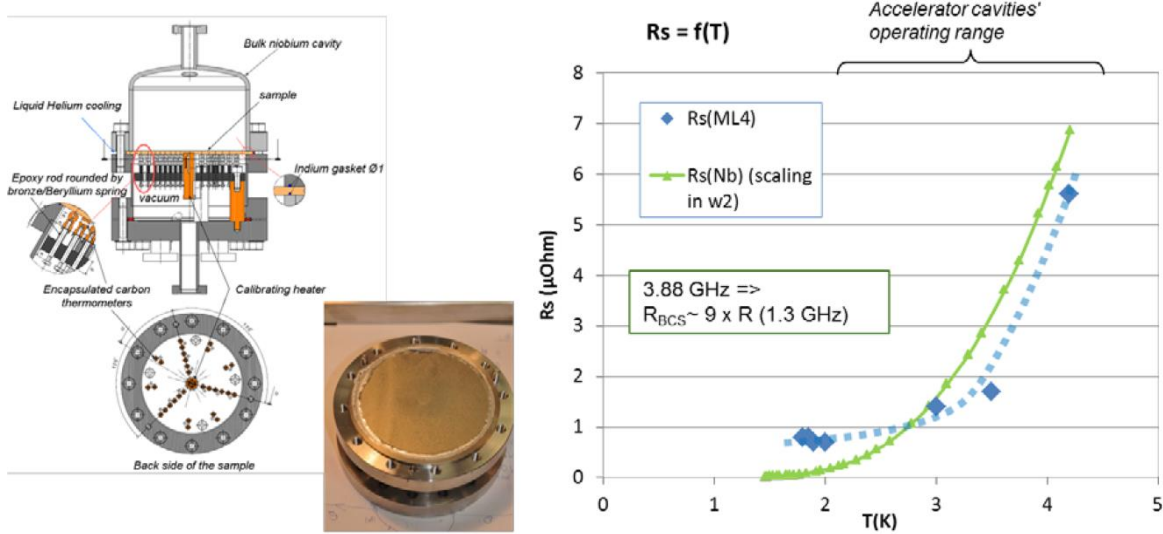

**Fig. 28:** The first RF test of a multilayer sample (4 nm × 25 nm of NbN). The samples constitute the removable bottom of a TE011 3.88 GHz cavity. The comparison is made with a 'good' bulk Nb cavity measured at 1.3 GHz and scaled to 3.88 GHz with an $\omega^2$ relationship. At low temperature, the residual resistance is dominant, but at this stage it is difficult to know its exact origin (indium seal or sample). At temperatures higher than 3 K (BCS regime), the multilayer sample exhibits a much lower surface resistance than pure Nb. In this configuration, half of the current flow inside the layers and the remaining half of the current is expected inside the niobium. Other

recent experimental measurements on similar films can be found in, for example, Ref. [93] (NbN) and Refs. [94, 95] (MgB$_2$).

**References**

[1]     R. Little and W. Whitney, "Electron emission preceding electrical breakdown in vacuum". Journal of Applied Physics, 1963. **34**(8): p. 2430-2432.

[2]     H. Padamsee, J. Knobloch, and T. Hays, "RF superconductivity for accelerators". 1998, 2nd Ed 2008: J. Wiley & son.

[3]     A. Descoeudres, et al., "DC breakdown conditioning and breakdown rate of metals and metallic alloys under ultrahigh vacuum". Physical Review Special Topics-Accelerators and Beams, 2009. **12**(3): p. 032001.

[4]     Werner. "Voltage breakdown and the processing mechanism". in SRF 2003. 2003. Tavemunde, Germany.

[5]     Werner. "Voltage breakdowns on Nb and Cu surfaces". in SRF 2001. 2001. Tsukuba, Japan.

[6]     A. Zeitoun-Fakiris and B. Juttner, "On the dose of bombarding residual gas ions for influencing pre-breakdown field emission in a vacuum,". J. Phys. D: Appl. Phys., 1991. **24**: p. 750-756.

[7]     Wang. "DC field emission studies on Nb". in SRF 2001. 2001. Tsukuba, Japan.

[8]     I. Brodie, "Studies of field emission and electrical breakdown between extended nickel surfaces in vacuum". Journal of Applied Physics, 1964. **35**(8): p. 2324-2332.

[9]     J. Tan, "Etude de l'émission électronique par effet de champ en haute fréquence". 1995, Paris-6, France.

[10]    G.R. Werner, "Probing and modeling voltage breakdown in vacuum". 2004, Cornell University.

[11]    E. Mahner, et al. "Field emission measurements on niobium cathodes of high purity". in SRF 91. 1991. DESY, Hamburg, Germany.

[12]    R. Noer, et al., "Electron field emission from intentionally introduced particles on extended niobium surfaces". Journal of Applied Physics, 1986. **59**(11): p. 3851-3860.

[13]    P. Niedermann, et al., "Field emission from broad area niobium cathodes: Effects of high temperature treatment". Journal of Applied Physics, 1986 **59**(3): p. 892-901.

[14]    T. Junquera, et al. "Study of luminous spots observed on metallic surfaces subjected to high RF fields". in PAC 95. 1995.

[15]    S. Maissa, "Etude des émissions lumineuses associées aux émissions électroniques dans les cavités hyperfréquences". 1996, Orsay Paris XI.

[16]    J.M. Jimenez, "Etude de l'émission électronique par effet de champ en courant continu sur des cathodes étendues". 1994, Clermont-II, France.

[17]    P. Niedermann and O. Fischer, "Application of a scanning tunneling microscope to field emission studies". IEEE Trans. on Dielectrics and Electrical Insulation, 1989 **24**(6): p. 905-915.

[18]    E.S. Crawford and J.S. Anderson, "Homogeneous solid state transformations in niobium oxides". Philosophical Transactions of the Royal Society of London A (Mathematical-and-Physical-Sciences), 1982. **304**(1485): p. 327-64.

[19]    J. Halbritter, "On the oxidation and on the superconductivity of niobium". Applied Physics A, 1987. **43**: p. 1-28.

[20]    M.L.A. Robinson and H. Roetschi, "AC polarisation in B-modification Nb$_2$O$_5$ single crystals". J. Phys. Chem. Solids, 1968. **29**: p. 1503-1510.

[21]    J.S. Sheasby and B. Cox, "Oxygen diffusion in Alpha-Niobium pentoxide". J. Less-Common Metals, 1968. **15**: p. 129-35.

[22] C. Chianelli, et al. "Very low current field electron emission from anodized niobium". in 5Th workshop on RF superconductivity. 1991.
[23] M. Jimenez, et al., "Electron field emission from selectively contaminated cathodes". Journal-of-Physics-D-(Applied-Physics), 1992. **26**(9): p. 1503-9.
[24] M. Jimenez and R. Noer, "Electron field emission from large-area cathodes: evidence for the projection model". J. Phys. D: Appl. Phys, 1994. **27**: p. 1038-1045.
[25] M. Luong, "Etude émission électronique par effet de champ sur des surfaces larges en régime statique et hyperfréquence". 1997, Universite Paris VI - Pierre et Marie Curie
[26] C.Z. Antoine, F. Peauger, and F. Le Pimpec, "Electromigration occurences and its effects on metallic surfaces submitted to high electromagnetic field: A novel approach to breakdown in accelerators". NIM A, 2012. **670**: p. 79-94.
[27] C.Z. Antoine. "Statistical analysis of the risk of dust contamination during assembling of RF cavities". in 6th Workshop on RF superconductivity. 1993. CEBAF, USA.
[28] J. Martignac, et al. "Particle contamination in vacuum systems". in 7th Workshop on RF Superconductivity. 1995. Gif sur Yvette,.
[29] H. Safa. "High field behavior of SCRF cavities". in 10th Workshop on RF Superconductivity. 2001. Tsukuba, Japan.
[30] H. Safa. "Ananalytical approach for calculating the quench field in superconducting cavities". in 7th workshop on RF superconductivity. 1995. Gif sur Yvette: DIST CEA.
[31] F. Koechlin and B. Bonin, "Parametrisation of the nobium thermal conductivity in the superconducting state.". Superconductor Science and Technology, 1996. **9**: p. 453-460.
[32] Issarovitch. "Development of centrifugal barrel polishing for treatment of superconducting cavities". 2003.
[33] T. Higuchi, et al. "Investigation on barrel polishing for superconducting niobium cavitie". in 7th SRF Internationnal Workshop. 1995. Gif-sur-Yvette, France.
[34] Saito. "Recent developments in SRF cavity cleaning techniques at KEK". in SRF 99. 1999. Santa Fe.
[35] C. Cooper. "Centrifugal Barrel Polishing of Cavities Worldwide". in SRF 2011. 2011. Chicago; Il, USA.
[36] C. Cooper and L. Cooley, "Mirror Smooth Superconducting RF Cavities by Mechanical Polishing with Minimal Acid Use". 2011, Fermi National Accelerator Laboratory (FNAL), Batavia, IL.
[37] Higuchi. "Development of hydrogen-free EP and hydrogen absorption phenomena". 2003.
[38] W.A. Shewhart, "Economic control of quality of manufactured product". Vol. 509. 1931: American Society for Qualit.
[39] C.Z. Antoine and R. Crooks, "Reducing Electropolishing Time with Chemical-Mechanical Polishing",SRF 2009, Berlin , 2009. http://accelconf.web.cern.ch/AccelConf/srf2009/html/author.htm
[40] A. Romanenko. "Review of High Field Q-slope, Surface Measurements". in SRF 2007. 2007. Bejing, China.
[41] A. Romanenko and H. Padamsee, "The role of near-surface dislocations in the high magnetic field performance of superconducting niobium cavities". Superconductor Science and Technology, 2010. **23**: p. 045008.
[42] O.S. Romanenko, "Surface Characterization Of Nb Cavity Sections-Understanding The High Field Q-Slope". 2009, Cornell University.
[43] A. Grassellino. "Muon Spin Rotation/Relaxation Studies of Niobium for SRF applications". in SRF 2011. 2011. Chicago; Il, USA.
[44] J. Zasadzinski, et al. "Raman spectroscopy as a probe of surface oxides and hydrides of niobium". in SRF 2011. 2011. Chicago, IL, USA: Thomas Jefferson National Accelerator Facility, Newport News, VA (United States).

[45] T. Proslier, et al., "Tunneling study of SRF cavity grade Nb: evidence of possible magnetic scattering at the surface.". APL, 2008. **92**: p. 212505.
[46] H.Shiba, Prog. Theor. Phys. 40, 435 (1968) et Prog. Theor. Phys. 50, 50 (1973).
[47] T. Proslier, et al., "Evidence of Surface Paramagnetism in Niobium and Consequences for the Superconducting Cavity Surface Impedance". Applied Superconductivity, IEEE Transactions on, 2011(99): p. 2619-2622.
[48] T. Proslier, et al., "Localized magnetism on the surface of niobium: experiments and theory". Bulletin of the American Physical Society, 2011. **56**.
[49] S. Casalbuoni, et al., "Surface superconductivity in Nb for superconducting RF cavities". Nuclear Instrumentation and Methods in Physical Research A, 2005. **538**: p. 45-64.
[50] M. Delheusy, "X-Ray investigation of Nb/Oxide interfaces". 2008, Orsay/Stuttgart: Orsay/Stuttgart.
[51] R.J. Cava, et al., "Electrical and magnetic properties of $Nb_{2}O_{5-d}$ crystallographic shear structures". PHYSICAL REVIEW B, 1991. **44**(13): p. 6973-6981.
[52] C. Antoine, "Materials and surface aspects in the development of SRF Niobium cavities". EUCARD series on Accelerator Science, ed. R.S. Romaniuk and J.P. Koutchouk. Vol. 12. 2012.
[53] F.P.-J. Lin and A. Gurevich, "Effect of impurities on the superheating field of Type II superconductors". Bulletin of the American Physical Society, 2012. **57**(1).
[54] M. Kharitonov, et al., "Surface impedance of superconductors with magnetic impurities". Arxiv preprint arXiv:1109.3395, 2011.
[55] J.P. Mercier, W. Kurz, and G. Zambelli, "Introduction à la science des matériaux". Vol. 1. 1999: PPUR.
[56] L.E. Samuels, "Metallographic polishing by mechanical methods". 2003: Asm Intl.
[57] S. Hashimoto, S. Miura, and T. Kubo, "Dislocation etch pits in gold". Journal of Materials Science, 1976. **11**(8): p. 1501-1508.
[58] X. Zhao, G. Ciovati, and T. Bieler, "Characterization of etch pits found on a large-grain bulk niobium superconducting radio-frequency resonant cavity". Physical Review Special Topics-Accelerators and Beams, 2010. **13**(12): p. 124701.
[59] J. Knobloch, et al. "High-Field Q Slope in superconducting cavities du to magnetic field enhancement at grain boundaries". in 9th Workshop on RF Superconductivity. 1999. Santa Fe , NM, USA.
[60] B. Visentin. "Review on Q-DROP mechanisms ". in International Workshop on Thin Films and New Ideas for Pushing the Limits of RF Superconductivity. 2006. Legnaro-Padue (Italie).
[61] H. Tian, et al., "A novel approach to characterizing the surface topography of niobium superconducting radio frequency (SRF) accelerator cavities". Applied Surface Science, 2010.
[62] C. Roques-Carmes, et al., "Geometrical description of surface topography by means of an equivalent conformal profile model.". Int. J. Mach. Tools Manufact., 1998. **38**(5-6): p. 573-579.
[63] M. Ge, et al., "Routine characterization of 3D profiles of SRF cavity defects using replica techniques". Superconductor Science and Technology, 2011. **24**: p. 035002.
[64] S. Berry, et al. "Topologic analysis of samples and cavities: a new tool for morphologic inspection of quench site". in 11th workshop on RF Superconductivity. 2003. Lübeck, Germany.
[65] S. Berry, C. Antoine, and M. Desmons. "Surface morphology at the quench site". in EPAC 2004. 2004. Lucern, Switzerland.
[66] A.A. Polyanskii, et al. "Review of magneto-optical result on high purity Nb for superconducting RF application.". in International workshop on thin films and new ideas for pushing the limit of RF superconductivity. 2006. Legnaro National Laboratories (Padua) ITALY.

[67] R. Ricker and G. Myneni, "Evaluation of the Propensity of Niobium to Absorb Hydrogen During Fabrication of Superconducting Radio Frequency Cavities for Particle Accelerators". Journal of Research of the National Institute of Standards and Technology, 2010. **115**(5).
[68] V. Shemelin and H. Padamsee, "Magnetic field enhancement at pits and bumps on the surface of superconducting cavities". TTC-Report, 2008. **7**: p. 2008.
[69] Singer. "Hydroforming of NbCu clad cavities at Desy". 2001.
[70] X. Singer, et al., "Hydroforming of Multi-Cell Niobium and NbCu-Clad Cavities". 2009, Thomas Jefferson National Accelerator Facility, Newport News, VA (United States).
[71] C.Z. Antoine, "RF material investigation by sample analysis". Particle accelerators, 1997. **60**(1-4).
[72] D. Landolt, "Fundamental aspects of electropolishing". Electrochimica Acta, 1987. **32**(1): p. 1-11.
[73] A. Valente-Feliciano, et al. "Development of Nb and Alternative Material Thin Films Tailored for SRF Applications". 2012.
[74] C. James, et al., "Superconducting Nb Thin Films on Cu for Applications in SRF Accelerators". Applied Superconductivity, IEEE Transactions on, 2013. **23**(3): p. 3500205-3500205.
[75] A. Anders, "Deposition of niobium and other superconducting materials with high power impulse magnetron sputtering: Concept and first results". 2013.
[76] T. Yogi, G. Dick, and J. Mercereau, "Critical rf magnetic fields for some type-I and type-II superconductors". Physical Review Letters, 1977. **39**(13): p. 826-829.
[77] E.H. Brandt. "Electrodynamics of superconductors exposed to high frequency fields". in International workshop on thin films and new ideas for pushing the limit of RF superconductivity. 2006. Legnaro National Laboratories (Padua) ITALY.
[78] A. Gurevich, "Multiscale mechanisms of SRF breakdown". Physica C, 2006. **441**(1-2): p. 38-43
[79] A. Gurevich and G. Ciovati, "Dynamics of vortex penetration, jumpwise instabilities, and nonlinear surface resistance of type-II superconductors in strong rf fields". Physical Review B 2008. **77**(10): p. 104501-21.
[80] P. Bauer, et al., "Evidence for non-linear BCS resistance in SRF cavities ". Physica C, 2006. **441**: p. 51–56.
[81] Saito. "Theoretical critical field in RF application". in SRF 03. 2003. Lubec.
[82] G. Lamura, et al., "First critical field measurements by third harmonic analysis ". Journal of Applied Physics, 2009. **106**: p. 053903
[83] see e.g. http://tdserver1.fnal.gov/project/workshops/RF_Materials/talks/Hasan_Sethna_superheating%20field.ppt
[84] H. Piel, "High T SC for accelerator cavities". NIM A, 1990. **287**(1-2): p. 294-309.
[85] http://tdserver1.fnal.gov/project/workshops/RF_Materials/talks/A-M_Valente-Feliciano_NewMaterialsOverview.ppt
[86] M. Fouaidy, et al. "New results on RF properties of superconducting niobium films using a thermometric system". in EPAC. 2002. Paris: European Physical Society.
[87] M. Mathur, D. Deis, and J. Gavaler, "Lower critical field measurements in NbN bulk and thin films". Journal of Applied Physics, 1972. **43**(7): p. 3158-3161.
[88] See e.g. presentations from Anders, Krishnan or Valente- Feliciano at thin film workshop 2010 or SRF 2011
[89] A. Gurevich, "Enhancement of RF breakdown field of SC by multilayer coating". Appl. Phys.Lett., 2006. **88**: p. 012511.
[90] C.Z. Antoine, J.C. Villegier, and G. Martinet, "Study of nanometric superconducting multilayers for magnetic field screening applications". Applied Physics Letters, 2013. **102**(10): p. 102603.

[91]   C.Z. Antoine, et al., "Characterization of field penetration in superconducting multilayers samples". IEEE Transactions on Applied Superconductivity, 2011. **21** (3): p. 2601 - 2604.

[92]   C.Z. Antoine, et al., "Characterization of superconducting nanometric multilayer samples for SRF applications: first evidence of magnetic screening effect". Physical Review Special Topics-Accelerators and Beams, 2010. **13**: p. 121001.

[93]   W. Roach, et al., "Magnetic Shielding Larger Than the Lower Critical Field of Niobium in Multilayers". Applied Superconductivity, IEEE Transactions on, 2013. **23**(3): p. 8600203-8600203.

[94]   T. Tajima and L.C. N.F. Haberkorn1, R.K. Schulze1, H. Inoue2, J. Guo3, V.A. Dolgashev3, D.W. Martin3, S.G. Tantawi3, C.G. Yoneda3, B.H. Moeckly4, T. Proslier5, M. Pellin5, A. Matsumoto6, E. Watanabe6, X.X. Xi7, C. Zhuang7, B. Xiao8. "MgB2 Thin Film Studies". in SRF 2011. 2011.

[95]   C. Zhuang, et al., "MgB2 Thin Films on Metal Substrates for Superconducting RF Cavity Applications". Journal of Superconductivity and Novel Magnetism, 2013: p. 1-6.